\documentclass[preprint2]{aastex}

\newcommand{\1}{{~\sc i}}
\newcommand{\2}{{~\sc ii}}
\newcommand{\3}{{~\sc iii}}
\newcommand{\4}{{~\sc iv}}
\newcommand{\5}{{~\sc v}}
\newcommand{\mic}{{\,$\mu$m}}

\usepackage{eucal} 

\shorttitle{Chemical composition and mixing in giant H\2\ regions}
\shortauthors{Lebouteiller et al.} 

\begin{document}


\title{Chemical composition and mixing in giant H\2\ regions: NGC\,3603, 30\,Doradus, and N\,66}


\author{V.\ Lebouteiller\altaffilmark{1}, J.\ Bernard-Salas\altaffilmark{1}, B.\ Brandl\altaffilmark{2}, D.G.\ Whelan\altaffilmark{1}, Yanling\ Wu\altaffilmark{1}, V.\ Charmandaris\altaffilmark{3}, D.\ Devost\altaffilmark{1}, J.R. Houck\altaffilmark{1}\altaffiltext{1}{Center for Radiophysics and Space Research, Cornell University, Space Sciences Building, Ithaca, NY 14853-6801, USA}\altaffiltext{2}{Leiden Observatory, University of Leiden, P.O. Box 9513, 2300 RA Leiden, Netherlands}\altaffiltext{3}{Department of Physics, University of Crete, GR-71003 Heraklion, Greece.; IESL/Foundation for Research and Technology-Hellas, GR-71110, Heraklion, Greece; and Chercheur Associ\'e, Observatoire de Paris, F-75014, Paris, France}}
\email{vianney@isc.astro.cornell.edu}






\begin{abstract}
We investigate the chemical abundances of NGC\,3603 in the Milky Way, of 30\,Doradus in the Large Magellanic Cloud, and 
of N\,66 in the Small Magellanic Cloud. 
Mid-infrared observations with the Infrared Spectrograph onboard the Spitzer Space Telescope allow us to probe the properties of distinct physical regions within each object: the central ionizing cluster, the surrounding ionized gas, photodissociation regions, and buried stellar clusters. 
We detect [S\3], [S\4], [Ar\3], [Ne\2], [Ne\3], [Fe\2], and [Fe\3] lines and derive the ionic abundances. 
Based on the ionic abundance ratio (Ne\3/H)/(S\3/H), we find that the gas observed in the MIR is characterized by a higher degree of ionization than the gas observed in the optical spectra. 
We compute the elemental abundances of Ne, S, Ar, and Fe. We find that the $\alpha$-elements Ne, S, and Ar scale with each other. 
Our determinations agree well with the abundances derived from the optical.
The Ne/S ratio is higher than the solar value in the three giant H\2\ regions and points toward a moderate depletion of sulfur on dust grains.
We find that the neon and sulfur abundances display a remarkably small dispersion (0.11\,dex in 15 positions in 30\,Doradus), suggesting a relatively homogeneous ISM, even though small-scale mixing cannot be ruled out.
\end{abstract}


\keywords{HII regions --- infrared: ISM --- ISM: atoms --- ISM: individual (NGC 3603, 30 Doradus, N 66)}

\section{Introduction}\label{sec:intro}

Giant H\2\ regions are ideal laboratories to understand the feedback of star-formation on the
dynamics and energetics of the interstellar medium (ISM). 
Supernov{\ae} and stellar winds arising in such regions are reponsible for producing shocks, destroying dust grains and molecules, while 
compressing molecular clouds and triggering subsequent star-formation. They also allow the release of newly synthetized elements into the ISM, altering its metallicity. 

In order to study the star-formation properties as a function of the environment, we observed three giant H\2\ regions spanning a wide range of physical conditions (gas density, mass, age) and chemical properties (metallicity) with the Spitzer Space Telescope (Werner et al.\ 2004).
Observations are part of the GTO program PID\#63. The regions are NGC\,3603 in the Milky Way, 30\,Doradus (hereafter 30\,Dor) in the Large Magellanic Cloud (LMC), and N\,66 in the Small Magellanic Cloud (SMC). The scope of this program is to address crucial issues such as the destruction of complex molecules by energetic photons arising from massive stars, the polycyclic aromatic hydrocarbon (PAH) abundance dependence on metallicity, or conditions that lead to the formation/disruption of massive stellar clusters.
Photometry with Spitzer/IRAC (Fazio et al.\ 2004) has been performed and will be discussed in Brandl et al.\ (in preparation).
The brightest mid-infrared (MIR) regions (knots, stellar clusters, shockfronts, ...) were followed spectroscopically with the Infrared Spectrograph (IRS; Houck et al.\ 2004). 
In Lebouteiller et al.\ (2007), we analyzed the spatial variations of the PAH and fine-structure line emission across individual photodissociation regions (PDRs) in NGC\,3603. 
The two other regions will be investigated the same way in follow-up papers (Bernard-Salas et al.\ in preparation; Whelan et al.\ in preparation). 
In this paper, we introduce the full IRS dataset (low- and high-resolution) of the giant H\2\ regions and we derive their chemical abundances. A subsequent paper will be focused on the study of molecules and dust properties (Lebouteiller et al.\ in preparation).

Elemental abundances in H\2\ regions are historically derived from optical emission-lines. Large optical telescopes, together with
sensitive detectors makes it possible to determine the chemical composition of very faint H\2\ regions. 
Because of dust extinction, optical spectra only observe ionized gas toward sighlines with low dust content. 
In this view, the MIR range allows analyzing denser lines of sight, with possibly different chemical properties because of 
small-scale mixing and/or differential depletion on dust grains. MIR emission-lines constitute the only way to measure 
abundances in more obscure regions, and these abundances ought to be compared to abundances from the optical range.
Although the optical domain gives access to some of the most important elements to constrain nucleosynthethic 
and stellar yields (C, N, O, Ne, S, Ar, Fe), it does not include some essential ionization stages necessary for abundance determinations of certain elements, such as S\4\ or Ne\2. 
The MIR range enables the abundance determination of Ne, S, and Ar, with the 
most important ionization stages observed. Iron abundance can be also determined from MIR forbidden emission-lines, but with considerably
larger uncertainty due to ionization corrections.
Finally, it must be stressed that abundance determinations in the optical are more sensitive to the electronic temperature ($T_e$) determination as compared to the MIR range. The effect of $T_e$ on abundances determinations is a significant source of error in optical abundance results.

Wu et al.\ (2007) recently studied a sample of blue compact dwarf galaxies (BCDs) with the IRS and found a global 
agreement between abundances derived from the optical and those derived from the MIR. 
This suggests that the dense lines of sight probed in the MIR have a similar chemical composition as unextincted lines of sight and/or 
dense regions with possibly peculiar abundances do not contribute significantly to the integrated MIR emission-line spectrum. 
MIR abundances of the BCDs were calculated using mostly H$\beta$ or H$\alpha$ lines from the optical as tracer of the hydrogen content, 
with significant uncertainties from aperture corrections, or different observed regions because of extinction. 
The present sample of giant H\2\ regions provides the unique opportunity to measure accurate abundances, with a signal-to-noise ratio
sufficiently high to observe directly the H\1\ recombination line at 12.37\mic. 
We provide abundances of Ne, S, Ar, and Fe toward lines of sight with different physical properties (PDRs, ionized gas, embedded source, stellar cluster, ...) within each giant H\2\ region.

We first present the sample of the three giant H\2\ regions in \S\ref{sec:pres}. The data reduction and analysis are discussed in \S\ref{sec:observations}. We infer the ion abundances in \S\ref{sec:ionicab}. Elemental abundances are determined in \S\ref{sec:eleab} and are discussed in \S\ref{sec:discussion}.

\section{Observations}\label{sec:pres}

\subsection{The sample of giant H\2\ regions}

\subsubsection{NGC\,3603}\label{sec:pres_3603}

\begin{figure*}
\includegraphics[angle=0,scale=0.7,clip=true]{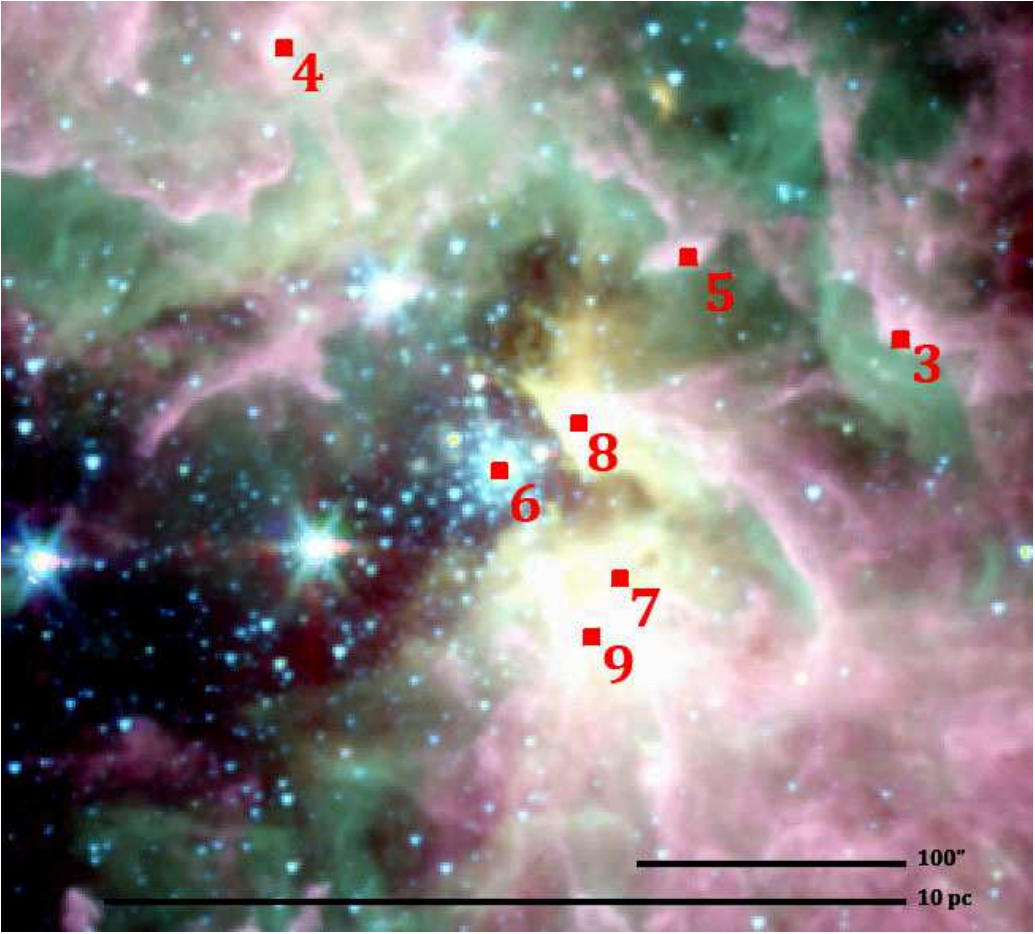}
\figcaption{IRS observations of NGC\,3603 overlaid on the preliminary IRAC image (Brandl et al.\ in preparation). We used the \textit{ch3} band ($[$5.0-6.4$]$\,$\mu$m) for the red, \textit{ch2} ($[$4.0-5.0$]$\,$\mu$m) for the green, and \textit{ch1} ($[$3.2-3.9$]$\,$\mu$m) for the blue. Positions \#1 and \#2 are located outside this field. \label{fig:IRAC3603}}
\end{figure*}

\begin{figure*}
\includegraphics[angle=0,scale=0.7,clip=true]{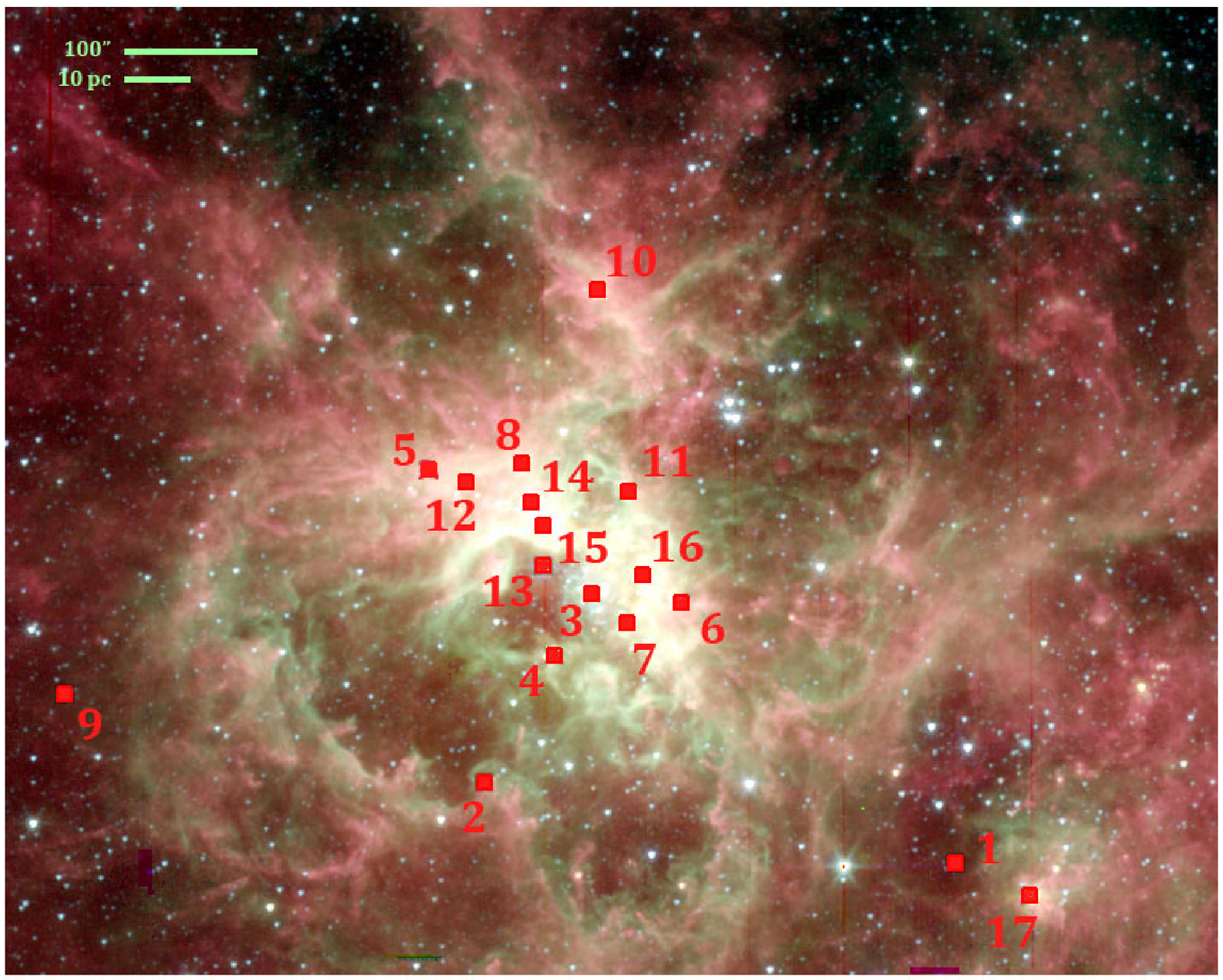}
\figcaption{IRS observations of 30\,Dor overlaid on the preliminary IRAC image (Brandl et al.\ in preparation). Colors are those defined in Fig.~\ref{fig:IRAC3603}. \label{fig:IRAC30dor}}
\end{figure*}

\begin{figure*}
\includegraphics[angle=0,scale=0.7,clip=true]{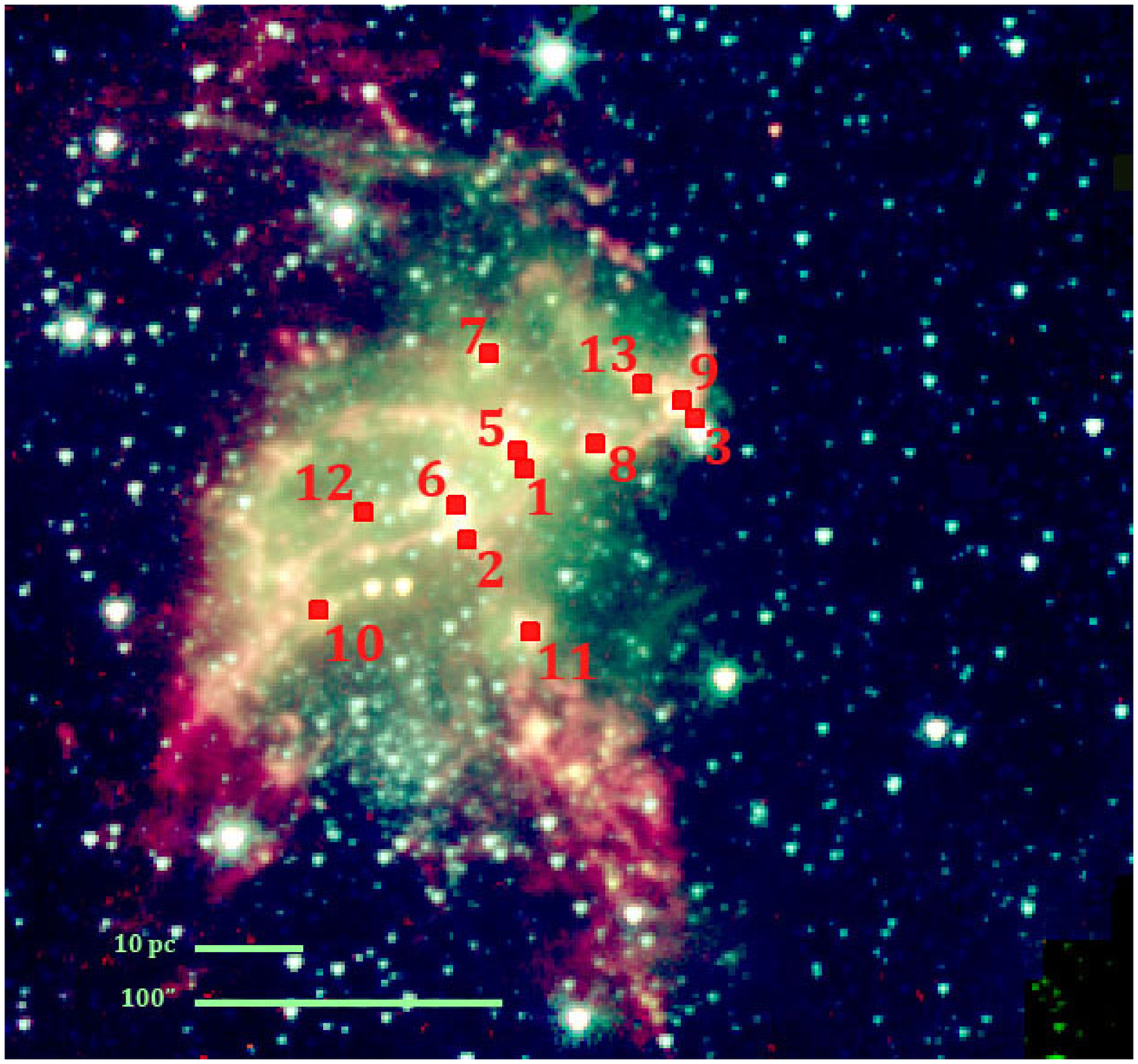}
\figcaption{IRS observations of N\,66 overlaid on the preliminary IRAC image (Brandl et al.\ in preparation). Colors are those defined in Fig.~\ref{fig:IRAC3603}. \label{fig:IRAC346}}
\end{figure*}

NGC\,3603 is located in the Carina arm of the Milky Way, at around $\approx$7\,kpc from the Sun (see e.g., Moffat et al.\ 1983; de Pree et al.\ 1999). Based on its Galactic longitude (291.62$^\circ$), NGC\,3603 is at around 8.5\,kpc from the Galactic Center, i.e., only somewhat further away than the Sun ($\approx$8\,kpc).
This is the most massive optically visible H\2\ region in the Milky Way (Goss \& Radhakrishnan 1969), being 100 times more luminous than the Orion Nebula.
NGC\,3603 is remarkably similar to R136, the core of 30\,Dor, in terms of its star density profile and its Wolf-Rayet (WR) content (Moffat et al.\ 1994). 
NGC\,3603 has essentially a solar metallicity with $12+\log$~(O/H)$\approx$8.39-8.52 (e.g., Melnick et al.\ 1989; Tapia et al.\ 2001; Garc{\'{\i}}a-Rojas et al.\ 2006; see also \S\ref{sec:eleab}). 

Massive stars in the central stellar cluster influence heavily the surrounding ISM morphology through stellar winds, notably by compressing molecular clouds (N{\"u}rnberger \& Stanke 2002). 
These massive stars are also responsible for most of the excitation in the H\2\ region through the large number of ionizing photons, with a Lyman continuum flux of $10^{51}$\,s$^{-1}$ (Kennicut 1984; Drissen et al.\ 1995).

\subsubsection{30\,Doradus}\label{sec:pres_30dor}

30\,Dor is located in the Large Magellanic Cloud, at a distance modulus of $(m-M)_0 = 18.45 \pm 0.15$ (Selman et al.\ 1999) i.e., $\approx49\pm3$\,kpc. 
It is the largest and most massive H\2\ region complex in the Local Group, with the nebula being $15'$ (200\,pc) in diameter.
The metallicity of 30\,Dor is a factor of $\sim$0.6 below solar (\S\ref{sec:eleab}).
The dense core of very luminous and massive stars is known as R136 ($\sim$2.5\,pc in diameter).
In R136, there are 39 confirmed O3 stars (Massey \& Hunter 1998) as well as several WR stars (e.g., Melnick 1985).
Stars from spectral types O3 to B3 are also detected as far away as 150$''$ from R136 (Bosch et al.\ 2001).

\subsubsection{N\,66 / NGC\,346}\label{sec:pres_346}

N\,66 is the largest nebula in the SMC (Henize 1956). 
The distance to N\,66 is about that of the SMC (60.6\,kpc; Hilditch et al.\ 2005).
Many massive stellar clusters are located across the region (Sabbi et al.\ 2007), but 
most of the nebular ionization is thought to be due to NGC\,346, the largest stellar concentration in the SMC (Dreyer 1888). 
The metallicity of the NGC\,346 cluster is $Z/$Z$_\odot=0.2\pm0.1$ (Haser et al.\ 1998; Bouret et al.\ 2003) and its age is $\sim$3\,Myr (Bouret et al.\ 2003). The other $-$ fainter $-$ clusters have similar ages (Sabbi et al.\ 2007).
Many H\2\ regions are located across the nebula, including the compact H\2\ region N66A, powered by its own stellar cluster.
Several dozen O stars are confirmed in NGC\,346, at least one of them as early as O3 type (Walborn \& Blades 1986; Massey et al.\ 1989).

\subsection{Observation strategy}\label{sec:obs}

\begin{deluxetable}{llllll}
\tabletypesize{\scriptsize}
\tablewidth{0pc}
\tablecolumns{4}
\tablecaption{Observation log.\label{tab:observations}}
\startdata
\tableline
\tableline
Pointing & $\alpha$ (J2000)  & $\delta$ (J2000)  & Campaign/AORKey & SH/LH\tablenotemark{a} & Spectral characteristics\tablenotemark{b} \\
\tableline
NGC3603\#1 & 11$^\textrm{h}$14$^\textrm{m}$30$^\textrm{s}$.00& -61$^\circ$19$'$40$''$.0 & 22/12080384& 0.25 & (Sky) \\
NGC3603\#3 & 11$^\textrm{h}$14$^\textrm{m}$49$^\textrm{s}$.06& -61$^\circ$17$'$09$''$.1& 22/12080384& 0.27 & PDR + SiO-abs \\
NGC3603\#4 & 11$^\textrm{h}$14$^\textrm{m}$56$^\textrm{s}$.71&-61$^\circ$12$'$56$''$.6& 22/12080384& 0.37 & PDR \\
NGC3603\#5 & 11$^\textrm{h}$14$^\textrm{m}$52$^\textrm{s}$.40&-61$^\circ$15$'$46$''$.3& 22/12080384& 0.32 & PDR + SiO-abs \\
NGC3603\#6 & 11$^\textrm{h}$15$^\textrm{m}$07$^\textrm{s}$.40&-61$^\circ$15$'$39$''$.2 &22/12080384& 0.26 & Stellar, SiO-em (HD97950)\\
NGC3603\#7 & 11$^\textrm{h}$15$^\textrm{m}$08$^\textrm{s}$.03&-61$^\circ$16$'$40$''$.2& 22/12080384& 0.27 & IG \\ 
NGC3603\#8 & 11$^\textrm{h}$15$^\textrm{m}$02$^\textrm{s}$.88&-61$^\circ$15$'$51$''$.6& 22/12080384& 1 & PDR + IG \\ 
NGC3603\#9 & 11$^\textrm{h}$15$^\textrm{m}$11$^\textrm{s}$.34&-61$^\circ$16$'$45$''$.2& 22/12080384& 0.44 & IG (IRS9A) \\ 
\tableline
30DOR\#1& 5$^\textrm{h}$38$^\textrm{m}$00$^\textrm{s}$.00&-69$^\circ$10$''$30$''$.0 &21.1/12081152& 0.02 & (Sky) \\
30DOR\#2& 5$^\textrm{h}$39$^\textrm{m}$03$^\textrm{s}$.80&-69$^\circ$08$'$06$''$.5 & 21.1/12081408& 0.29 & PDR \\
30DOR\#3& 5$^\textrm{h}$38$^\textrm{m}$42$^\textrm{s}$.43&-69$^\circ$06$'$02$''$.2 & 21.1/12081408& 0.28 & PDR + IG (R136) \\
30DOR\#4& 5$^\textrm{h}$38$^\textrm{m}$49$^\textrm{s}$.76&-69$^\circ$06$'$43$''$.1 & 21.1/12081408& 0.375 & IG \\
30DOR\#5& 5$^\textrm{h}$39$^\textrm{m}$01$^\textrm{s}$.30&-69$^\circ$04$'$00$''$.5&  21.1/12081408& 0.28 & PDR \\
30DOR\#6& 5$^\textrm{h}$38$^\textrm{m}$30$^\textrm{s}$.13& -69$^\circ$06$'$25$''$.1&  21.1/12081408& 0.24 & PDR \\
30DOR\#7& 5$^\textrm{h}$38$^\textrm{m}$38$^\textrm{s}$.44& -69$^\circ$06$'$31$''$.0&  21.1/12081408 &0.31 & PDR + IG (P644) \\
30DOR\#8& 5$^\textrm{h}$38$^\textrm{m}$48$^\textrm{s}$.11&-69$^\circ$04$'$12$''$.2 & 21.1/12081408 & 0.36 & PDR (P4\tablenotemark{c} )\\
30DOR\#10& 5$^\textrm{h}$38$^\textrm{m}$31$^\textrm{s}$.58&-69$^\circ$02$'$14.$''$4 & 21.1/12081152 &0.60 & PDR \\
30DOR\#11& 5$^\textrm{h}$38$^\textrm{m}$34$^\textrm{s}$.01&-69$^\circ$04$'$52$''$.1 & 21.1/12081152& 0.39 & PDR + IG  \\
30DOR\#12& 5$^\textrm{h}$38$^\textrm{m}$56$^\textrm{s}$.47&-69$^\circ$04$'$16$''$.7 & 21.1/12081152& 0.37 & PDR + SiO-abs \\
30DOR\#13& 5$^\textrm{h}$38$^\textrm{m}$48$^\textrm{s}$.41&-69$^\circ$05$'$32$''$.9 &  21.1/12081152& 0.29 & SiO-em  \\
30DOR\#14& 5$^\textrm{h}$38$^\textrm{m}$47$^\textrm{s}$.99&-69$^\circ$04$'$42$''$.9 & 21.1/12081152& 0.35 & PDR (P1429) \\
30DOR\#15& 5$^\textrm{h}$38$^\textrm{m}$46$^\textrm{s}$.93&-69$^\circ$05$'$02$''$.5 &21.1/12081152&   0.31 & PDR (P1\tablenotemark{c} ) \\
30DOR\#16& 5$^\textrm{h}$38$^\textrm{m}$34$^\textrm{s}$.58&-69$^\circ$05$'$57$''$.5 & 21.1/12081152&  0.32 & IG (P3\tablenotemark{c} ) \\
30DOR\#17& 5$^\textrm{h}$37$^\textrm{m}$50$^\textrm{s}$.36&-69$^\circ$11$'$07$''$.1 & 21.1/12081152&   0.65 & SiO-abs (2MASS J05375004-6911075) \\
\tableline
N66\#1 & 0$^\textrm{h}$59$^\textrm{m}$09$^\textrm{s}$.91&-72$^\circ$10$'$51$''$.0 &8/4384768& 0.18 &PDR + IG \\
N66\#2 & 0$^\textrm{h}$59$^\textrm{m}$06$^\textrm{s}$.63&-72$^\circ$10$'$25$''$.0 &8/438476& 0.12  &PDR \\
N66\#3 & 0$^\textrm{h}$59$^\textrm{m}$21$^\textrm{s}$.52&-72$^\circ$11$'$17$''$.1 &8/438476 & 0.25 & PDR\\
N66\#5 & 0$^\textrm{h}$59$^\textrm{m}$09$^\textrm{s}$.24&-72$^\circ$10$'$56$''$.9 &26/16207872& 1.06 & SiO-abs \\
N66\#6& 0$^\textrm{h}$59$^\textrm{m}$05$^\textrm{s}$.52&-72$^\circ$10$'$35$''$.9 & 26/16207872& 1.23 & SiO-em (NGC\,346)\\
N66\#7& 0$^\textrm{h}$59$^\textrm{m}$05$^\textrm{s}$.97&-72$^\circ$11$'$26$''$.9 &26/16207872& 1 & PDR + IG \\
N66\#8& 0$^\textrm{h}$59$^\textrm{m}$14$^\textrm{s}$.68&-72$^\circ$11$'$03$''$.2 &26/16207872 &0.75 & PDR \\
N66\#9& 0$^\textrm{h}$59$^\textrm{m}$20$^\textrm{s}$.43&-72$^\circ$11$'$22$''$.1 &26/16207872& 0.89 & PDR \\
N66\#10& 0$^\textrm{h}$58$^\textrm{m}$56$^\textrm{s}$.95&-72$^\circ$09$'$54$''$.0 & 26/16207872& 1 & PDR \\
N66\#11& 0$^\textrm{h}$59$^\textrm{m}$12$^\textrm{s}$.29&-72$^\circ$09$'$58$''$.2 & 26/16207872& 1 & PDR + IG + SiO-em \\
N66\#12& 0$^\textrm{h}$58$^\textrm{m}$59$^\textrm{s}$.02&-72$^\circ$10$'$28$''$.5 & 26/16207872& 0.54 & PDR + IG \\
N66\#13& 0$^\textrm{h}$59$^\textrm{m}$17$^\textrm{s}$.30&-72$^\circ$11$'$25$''$.2 & 26/16207872& 1 & PDR \\
\tableline
\enddata
\tablenotetext{a}{Scaling factor applied to to LH module spectra.}
\tablenotetext{b}{Dominant spectral characteristics of the MIR spectrum, with 'PDR' for strong PAH band emission, 'IG' for ionized gas with prominent [S\4] and [Ne\3] lines, 'SiO-abs' for the presence of silicate absorption, 'SiO-em' for the presence of silicate emission , and 'Stellar' for the presence of a stellar continuum emission.}
\tablenotetext{c}{ID in the infrared source list of Hyland et al.\ (1992).}
\end{deluxetable}

An observation log is presented in Table~\ref{tab:observations} where we report the coordinates of each position, the module scaling factor (\S\ref{sec:stitching}), and the spectral characteristics. 
A total of seven positions were observed in NGC\,3603 (Fig.~\ref{fig:IRAC3603}), fifteen positions in 30\,Dor (Fig.~\ref{fig:IRAC30dor}), and twelve positions in N\,66 (Fig.~\ref{fig:IRAC346}).
We focus our discussion on the observations from the SH and LH modules, which cover 9.9-19.6\mic\ and 18.7-37.2\mic\ respectively, with a spectral resolution
of $R\sim600$. Observations from the SL module (5.2-14.5\mic; $R\sim60-127$) were used to extend the spectral coverage shortward of 9.9\mic. 

We also included the first IRS observation of N\,66, originally designed to probe MIR bright knots as part of PID\#63, 
but which resulted in a mispointing. 
These observations (N66\#1, N66\#2, and N66\#3; Table~\ref{tab:observations}) probe relatively low-excitation regions, with a few arcseconds offset from the originally intended MIR sources.

\section{Data reduction and analysis}\label{sec:observations}

\subsection{Image cleaning and reduction}

The two-dimensional detector images were processed by the Spitzer Science Center's pipeline reduction
software (version S13.2). We used the the basic calibrated data product.
Rogue pixels and on-the-fly flagged data were removed using IRSCLEAN\footnote{The IRSCLEAN package can
be downloaded from \textit{http://ssc.spitzer.caltech.edu}}.

\subsection{Sky substraction}

No sky-subtraction was performed for 30\,Dor and N\,66 since the observations initially designed to 
be background images include prominent MIR emission features (lines and PAHs). It must be noted that the 
regions are bright enough that the lack of background subtraction does not affect the measurements.
We took instead the opportunity to use these background spectra and investigate low-excitation regions.
Source \#1 was used for sky-subtraction in NGC\,3603; it shows extremely weak emission-lines, with fluxes less than 1\% that of lines
in other positions.

\subsection{Extraction}

High-resolution spectra were extracted from the full SH and LH apertures using the SMART software\footnote{The SMART software can be downloaded from \textit{http://isc.astro.cornell.edu/Main/SmartRelease}} developped at Cornell (Higdon et al. 2004).
The low-resolution SL spectra were taken from Lebouteiller et al.\ (2007) for NGC\,3603, Bernard-Salas et al.\ (in preparation) for 30\,Dor, and Whelan
et al.\ (in preparation) for N\,66.

\subsection{Order stitching and source extent}\label{sec:stitching}

The spectra from the SH and LH modules had to be scaled for most observations because the aperture sizes are different and 
because the sources sampled are not unresolved. The SH aperture covers $4.7''\times11.3''$ on the sky while the LH one covers \ $11.1''\times22.3''$ and the two apertures are perpendicular to each other. 
If these apertures were uniformly illuminated, this would result in a correction factor of $\sim0.215$ in the ratio of the SH to LH flux. 

We applied a scaling factor to the flux in the LH observations to scale to the SH ones. Note that we align the dust 
continuum, and that the lines might behave differently (see the discussion in \S\ref{sec:density}).
Correction factors are reported in Table\,\ref{tab:observations} and range from $\sim$0.25 to 1. 
The only source for which the factor is significantly greater than 1 is N66\#6. The SH observation contains an additional source as compared to the LH one, which explains the additional flux observed in the SH spectrum.
For several other sources in N\,66, no correction was needed, implying that the extent of those MIR sources is smaller than the 
SH aperture. 
The MIR sources in NGC\,3603 and 30\,Dor show SH/LH scaling factors usually between $\sim$0.3 and $\sim$0.4, i.e., consistent with extended emission.

The distance to 30\,Dor and N\,66 is similar, but these objects show strong differences in the extent of their MIR sources.
This suggests that the MIR emission in N\,66 is mostly concentrated in small knots and that there is no intense extended dust continuum emitting across the whole nebula (see also Whelan et al.\ in preparation).

We also aligned spectra from the SL module to stitch those from the SH/LH modules.
Final spectra are presented in the appendix.

\subsection{Measurements}\label{sec:measurements}

\begin{figure}[h!]
\includegraphics[angle=0,scale=0.45,clip=true]{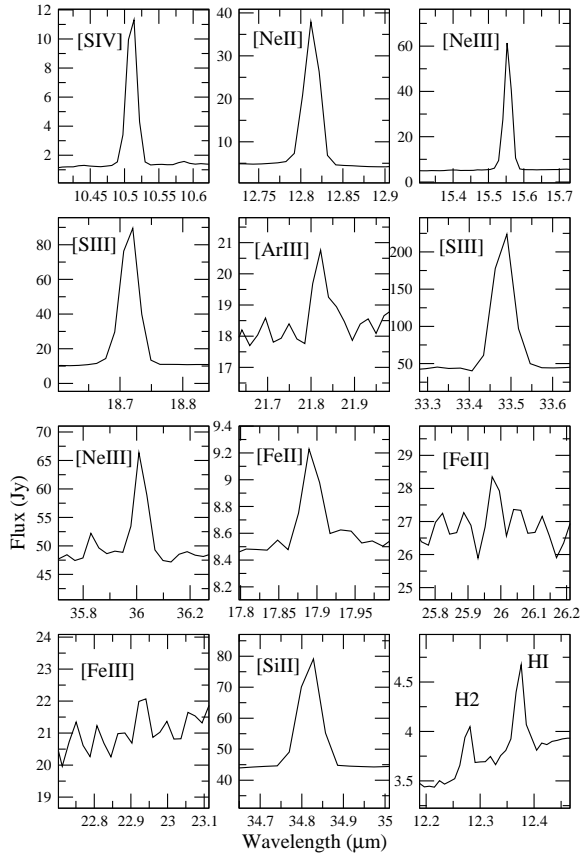}
\figcaption{Detection of the most important lines in the observation 30DOR\#8.
\label{fig:lines}}
\end{figure}

We illustrate in Fig.~\ref{fig:lines} the detection of the most important lines in the MIR spectrum of one
observation. 
Line fluxes are presented in Table~\ref{tab:line_fluxes}.
In order to infer the flux from a given emission line, we adjusted a first-order continuum to fit the data on
both sides of the line. The line was then fitted with a gaussian profile. Measurements were done in the 
combined data where the spectral orders and the two nod positions have been co-added.

\begin{deluxetable}{llllllllllllll}
\tablecolumns{13}
\tablewidth{0pc}
\tabletypesize{\scriptsize}
\tablecaption{Line fluxes.\label{tab:line_fluxes}}
\startdata
\tableline
\tableline
Line              &  [S\4]  & [Ne\2]   & [Ne\3]  & [S\3]   & [Ar\3]  & [S\3]   & [Ne\3]  & [Fe\2]  & [Fe\2]  & [Fe\3]  & [Si\2]  & H\1\   \\
$\lambda$ (\mic)  &  10.51 & 12.81 & 15.56 & 18.71 & 21.83 & 33.48 & 36.01 & 17.89 & 26.00 & 22.93 & 34.82 & 12.37 \\
\tableline
NGC3603\#3       &  53.83 & 	134.45 & 	181.10 & 	142.69 & 	5.60 &  	291.98 & 	24.35 & 	1.94 & 	3.05 & 	3.23 & 	 55.33 &     3.50   \\
NGC3603\#4       & 49.30 & 	280.13 & 	162.93 & 	399.72 & 	7.82 & 	        542.34 & 	29.58 & 	2.48 & 	8.33 & 	9.55 & 	 137.78 &    5.90   \\
NGC3603\#5       & 83.81 & 	89.73 & 	193.42 & 	209.58 & 	6.27 & 	        304.99 & 	28.50 & 	1.23 & 	3.66 & 	3.30 & 	 58.82 &     2.98   \\
NGC3603\#6       & 20.98 & 	11.77 & 	36.80 & 	35.40 & 	1.45 &    	52.25 & 	6.85 & 	$<$0.42 & 	$<$1.34 & 	0.86 & 	 8.94 & 	  5.86:\tablenotemark{a}   \\
NGC3603\#7       & 1075.34 &      140.92 & 	\nodata\tablenotemark{b} &          833.37 & 	\nodata\tablenotemark{b} & 	\nodata\tablenotemark{b} & 	134.64 & 	$<$7.75 & \nodata\tablenotemark{b} &   \nodata\tablenotemark{b}  &  $<$24.38 &    14.07  \\ 
NGC3603\#8       & 1725.83  &     429.63 &      \nodata\tablenotemark{b}	 &   \nodata\tablenotemark{b}	 & 	    \nodata\tablenotemark{b}    & 	\nodata\tablenotemark{b} & 	541.07 & 	$<$16.7 &  \nodata\tablenotemark{b}&  \nodata\tablenotemark{b}  &   $<$131.23 &   21.32  \\
NGC3603\#9       & 521.82 & 	246.0 &          \nodata\tablenotemark{b}	& 	        635.09 & \nodata\tablenotemark{b} & 	\nodata\tablenotemark{b} & 	153.30 & 	$<$11.73  & \nodata\tablenotemark{b} & 	\nodata\tablenotemark{b}  &  $<$96.26  &   6.63:\tablenotemark{a}   \\
\tableline
 30DOR\#2     & 4.35&	3.18&	12.50&	9.83&	0.26&	16.80&	1.22&	0.08&	0.07&	0.19&	 2.95&	  0.24    \\ 
  30DOR\#3     & 36.12&	8.00&	60.19&	37.77&	0.94&	59.25&	5.73&	0.31&	0.18&	0.60&	 7.16&	  1.00    \\ 
 30DOR\#4     & 102.71&	7.84&	107.02&	53.26&	1.52&	68.58&	8.68&	0.58&	$<$0.50&	1.29&	 3.43&	  1.56    \\ 
30DOR\#5     & 8.63&	5.94&	20.99&	17.34&	0.51&	31.71&	2.44&	0.28&	0.37&	0.39&	 9.46&	  0.37    \\ 
 30DOR\#6     & 15.67&	7.08&	33.33&	25.05&	0.51&	41.45&	3.69&	0.19&	0.86&	0.84&	 8.92&	  0.50    \\ 
 30DOR\#7     & 105.95&	15.05&	143.96&	79.73&	2.59&	109.26&	12.39&	0.58&	0.92&	1.91&	 8.85&	  2.37    \\ 
 30DOR\#8     & 18.31&	34.39&	62.07&  64.08&	2.05&	111.14&	7.10&	0.62&	1.93&	2.02&	 24.33 &   1.74    \\ 
 30DOR\#10     & 13.6&	39.57&	31.68&	54.29&	1.50&	66.96&	3.74&	0.37&	1.53&	1.47&	 19.63 &   1.40    \\ 
 30DOR\#11     & 32.39&	14.35&	74.99&	54.82&	1.63&	98.25&	8.45&	0.54&	0.84&	1.33&	 11.72 &   1.15    \\ 
 30DOR\#12     & 31.75&	6.26&	46.80&	28.02&	0.98&	64.03&	6.26&	0.34&	0.60&	0.68&	 11.93&    0.66    \\ 
 30DOR\#13     & 7.63&	2.61&	14.90&	10.73&	0.35&	22.87&	2.66&	0.10&	$<$0.15&	0.29&	 3.49&	  0.27    \\ 
 30DOR\#14     & 131.11&	27.69&	207.41& 117.35&	3.14&	168.87&	15.93&	0.86&	2.75&	2.54&	 34.57 &   3.09    \\ 
 30DOR\#15     &  151.33&	28.29&	245.48&	142.31&	4.55&	222.20&	22.61&	0.83&	2.40&	3.10&	 24.24 &   3.13    \\ 
 30DOR\#16     & 150.39&	14.23&	183.24&	97.64&	2.92&	126.17&	14.16&	0.81&	$<$1.41&	1.95&	 10.66 &   2.78    \\ 
 30DOR\#17     & 1.74&	7.75&	7.08&	10.18&	0.53&	34.65&	1.58&	0.20&	1.65&	1.07&	 13.20&    0.32    \\ 
\tableline
 N66\#1    &3.06&	0.31&	3.24&	2.24&	0.05&	2.03&	0.25&	0.26&	$<$0.05&	0.01&	 0.54&	  0.09       \\ 
 N66\#2    &2.31&	0.28&	2.41&	1.64&	0.03&	1.03&	0.13&	0.11&	$<$0.04	&$<$0.03&	 0.37&	  0.083      \\ 
 N66\#3    &1.46&	0.12&	1.47&	0.95&	0.04&	1.18&	0.16&	0.04&	$<$0.04&	$<$0.03	& 0.31&	 $<$0.01       \\ 
 N66\#5    &2.78&	0.42&	3.59&	2.41&	0.13&	6.25&	0.64&	0.16&	$<$0.08&	$<$0.10	& 1.40&	  0.15       \\ 
 N66\#6    &2.53&	0.69&	2.65&	2.11&	0.15&	5.64&	0.64&	$<$0.11&	0.24&	$<$0.14&	 2.54&	  0.10       \\ 
 N66\#7    &1.73&	0.18&	2.12&	1.43&	0.12&	3.90&	0.52&	$<$0.11&	0.22&	$<$0.14&	 2.52&	 0.13      \\  
 N66\#8    &3.09&	0.88&	4.49&	3.32&	0.10&	4.05&	0.39&	$<$0.06&	0.07&	0.04&	 0.84&	  0.16       \\ 
 N66\#9    &1.49&	0.55&	2.53&	1.89&	0.07&	3.25&	0.44&	$<$0.07&	$<$0.07&	$<$0.08	& 1.28&	  0.09       \\ 
 N66\#10    &1.50&	0.55&	2.52&	1.87&	0.07&	3.25&	0.52&	$<$0.07&	$<$0.07&	$<$0.08&	 1.29&	  0.10     \\   
 N66\#11   &1.49&	0.55&	2.51&	1.94&	0.07&	3.27&	0.29&	$<$0.07&	$<$0.07&	$<$0.08&	 1.25&	  0.08     \\   
 N66\#12    &3.54&	0.26&	3.11&	1.65&	0.09&	2.30&	0.32&	$<$0.10&	$<$0.05&	$<$0.04&	 0.48&	  0.13     \\   
 N66\#13    &2.20&	0.38&	2.88&	2.04&	0.08&	4.36&	0.42&	$<$0.10&	$<$0.07&	$<$0.08&	 0.97&	  0.12     \\   
\tableline
\enddata
\tablecomments{In units of $\times10^{-20}$\,W\,cm$^{-2}$. We estimate the errors on the line fluxes to be 10\% for [S\3], [S\4], [Ne\2], [Ne\3], [Si\2], [Fe\2] and H\1, and 15\% for [Ar\3] and [Fe\3] (\S\ref{sec:measurements}).}
\tablenotetext{a}{Blended with a broad stellar component (see \S\ref{sec:measurements}).}
\tablenotetext{b}{Saturated.}
\end{deluxetable}

The [Ne\2] line at 12.81\mic\ is blended with a PAH feature at $\sim$12.7\mic,
and for this particular case we forced the continuum to fit the data around 12.75\mic. The observations NGC3603\#6 and NGC3603\#9 show a broad emission bump at $\sim$12.4\mic\ $-$ seen in both low-resolution
and high-resolution spectra $-$ which is possibly stellar in origin. 
The 12.28\mic\ H$_2$ line and the 12.37\mic\ H\1\ line are blended with this broad feature, and their contribution to the integrated emission 
is not clearly visible. For this reason, it was not possible to reliably measure their flux.  

The statistical errors from the fit are smaller than other uncertainties such as errors on the flux calibration and stitching of the SH and LH module spectra (\S\ref{sec:stitching}). We estimate the total measurement errors to range from $\sim$10\% (for H\1, [Ne\2], [Ne\3], [Si\2], [S\3], [S\4], and [Fe\2]) to $\sim$20\% (for [Ar\3] and [Fe\3]).

\section{Ionic abundances}\label{sec:ionicab}

\subsection{Method}\label{sec:ionicab_method}

Fine-structure lines can be used to measure the ionic abundances relative to hydrogen. 
To derive the ionized hydrogen content, we made use of the H\1\ recombination line at 12.372\mic\ 
(Humphreys~$\alpha$, 7$\rightarrow$6). The Hu$\alpha$ line
is detected in all the positions of each object, except in N66\#3, where we calculated an upper limit (Table~\ref{tab:line_fluxes}). The positions NGC3603\#6 and NGC3603\#9 show a broad stellar emission bump preventing a reliable H\1\ line flux estimate (\S\ref{sec:measurements}).
For all the other positions, the Hu$\alpha$ line is blended with another, relatively weaker, H\1\ line at 12.387\mic\ (11$\rightarrow$8 transition). We estimated and corrected for
the  contribution of this line using the tables of Hummer \& Storey (1987) for case B recombination to calculate the flux ratio H\1~$_{\rm 11-8}$/H\1~$_{\rm 7-6}$ under the physical conditions described in \S\ref{sec:density}. 
The contribution of the H\1~$_{\rm 11-8}$ line that we corrected for is $\sim12\%$.

To determine the ionic abundance, we first estimate the H$\beta$ flux from the Hu$\alpha$ line using the tables of Hummer \& Storey (1987) for a given electron density and temperature. We then infer the ionic abundances using the method described in Bernard-Salas et al.\ (2001). The ionic abundance is defined as:

\begin{equation}
\frac{ N_{\rm ion} }{ N_{\rm p} } = \frac{ I_{\rm ion} }{ I_{\rm p} } N_e \frac{ \lambda_{\rm ul} }{ \lambda_{\rm H\beta} } \frac{ \alpha_{\rm H\beta} }{ A_{\rm ul} } \left(\frac{ N_{\rm u} }{ N_{\rm ion} }\right)^{-1},
\end{equation}

where $N_{\rm p}$ is the density of protons, $I_{\rm ion}$/~$I_{\rm p}$ is the ratio of observed intensities, $\lambda_{\rm ul}$ is the wavelength of the line and 
$\lambda_{\rm H\beta}$ is the wavelength of H$\beta$, $\alpha_{\rm H\beta}$ is the effective recombination coefficient for H$\beta$, A$_{\rm ul}$ is the Einstein spontaneous transition rate for the line and finally $N_{\rm u}$/ $N_{\rm ion}$ is the ratio of the population of the level from which the line 
originates to the total population of the ion. This ratio is obtained by solving the statistical equilibrium equation for a five level system and normalizing the total number of ions (Osterbrock 1989). Collisional strengths were taken from the IRON project\footnote{Find the IRON project at \textit{http://www.astronomy.ohio-state.edu/$\sim$pradhan/ip.html}.} (Hummer et al.\ 1993).

\subsection{Physical conditions}\label{sec:density}

\begin{figure}[t!]
\includegraphics[angle=0,scale=0.45,clip=true]{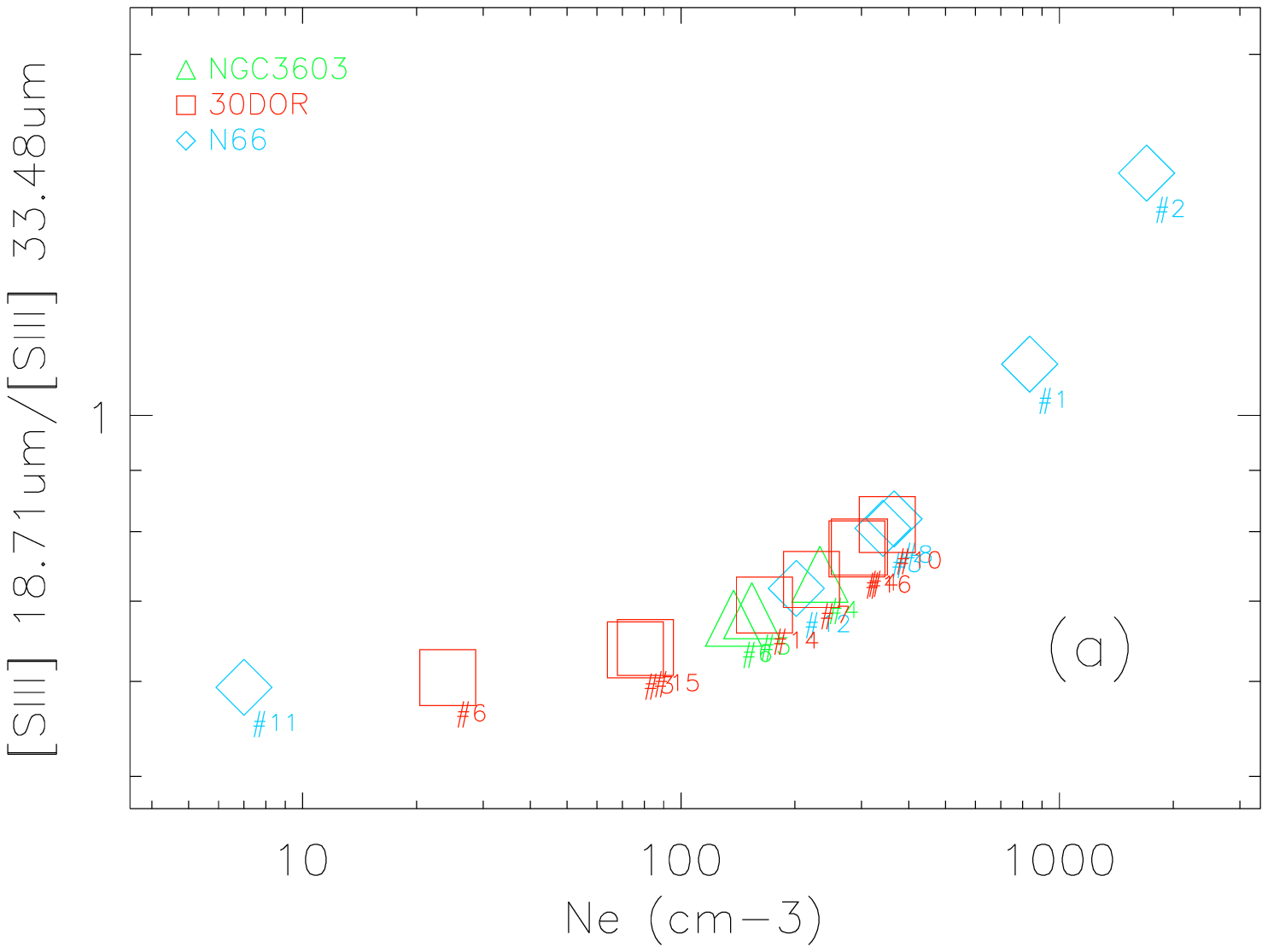}\\
\includegraphics[angle=0,scale=0.45,clip=true]{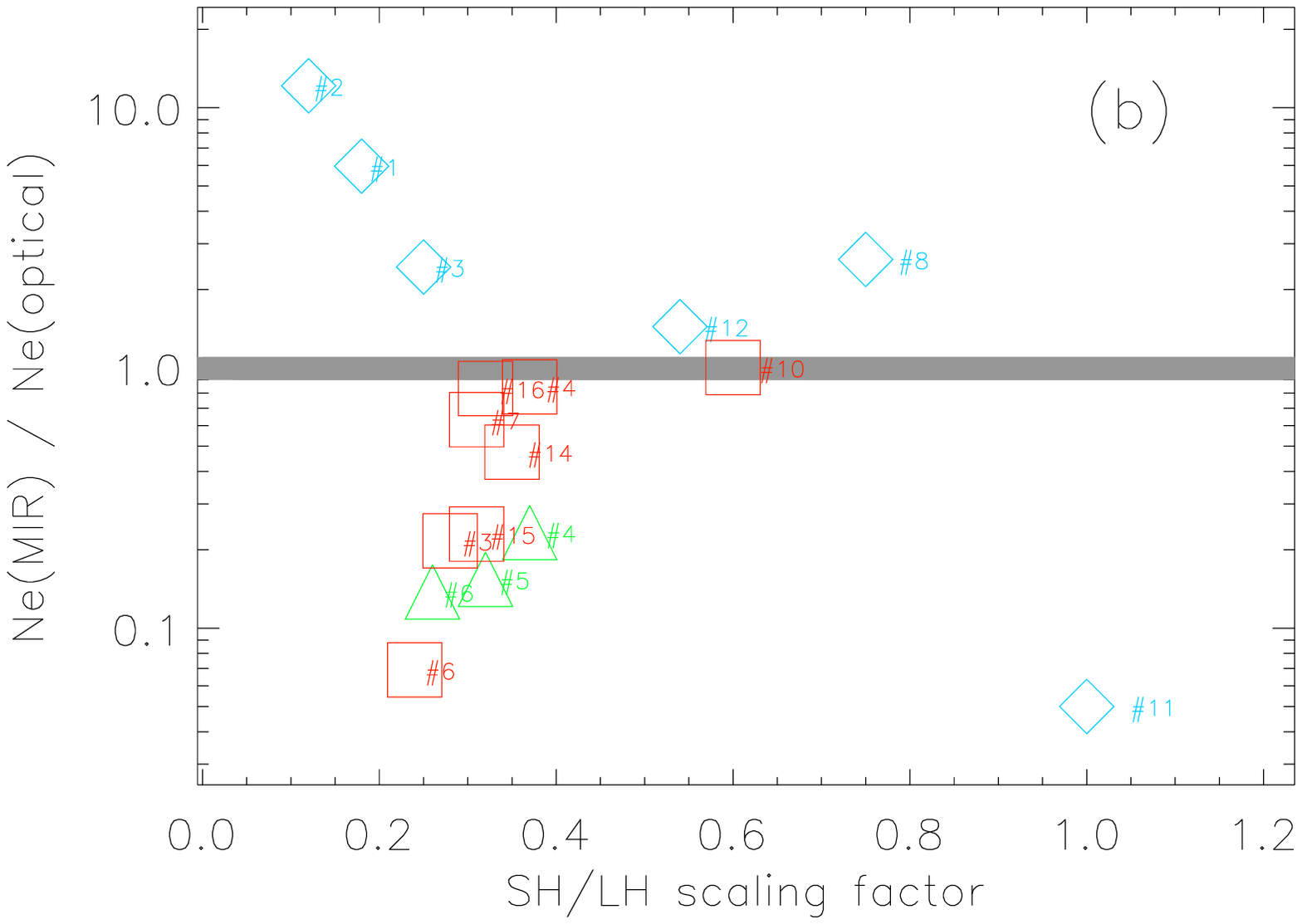}
\figcaption{\textit{Top - }The [S\3] line ratio is plotted as a function of the electron density. \textit{Bottom - } The ratio of the electron density measured from the MIR and optical is plotted as a function of the module scaling factor (see Table~\ref{tab:observations}). For the optical references, we use $N_e=1000$\,cm$^{-3}$ in NGC\,3603 (Garcia-Rojas et al.\ 2006; Simpson et al.\ 1995), $N_e=350$\,cm$^{-3}$ in 30\,Dor (Peimbert 2003; Vermeij \& ven der Hulst 2002), and $N_e=140$\,cm$^{-3}$ in N\,66 (Peimbert et al.\ 2000).
\label{fig:density}}
\end{figure}

In order to calculate the ionic abundances, it is necessary to estimate the electron density ($N_e$) and the temperature ($T_e$) in the nebula.
It is possible to estimate the electron density notably through diagnostics provided by the measurements of different lines from the same ionization stage of a given ion.
By combining SH and LH observations, we have access to two independent line ratios: [S\3]$_{18.71{\mu}m}$/[S\3]$_{33.48{\mu}m}$, and [Ne\3]$_{15.56{\mu}m}$/[Ne\3]$_{36.01{\mu}m}$.
The [Ne\3] line ratio is irrelevant for our data because it is sensitive to much higher densities ($N_e\gtrsim10^4$\,cm$^{-3}$) than what is usually found in giant H\2\ regions.

The [S\3] line ratio is sensitive to lower densities, but most of our data points lie in the flat regime where no reliable
density can be determined (Fig.~\ref{fig:density}a). 
It must be stressed that the [S\3] lines from which we derive the density do not
belong in the same module. Hence the [S\3] line ratio determination is strongly affected by uncertainties in the module scaling factor. 
Although the spectra were stitched to align (\S\ref{sec:stitching}), the regions observed in the two modules are not necessarily the same and might not share similar physical conditions. We aligned the SH and LH spectra based on the dust continuum and it is likely that the line fluxes do not scale accordingly.
In fact, it can be seen in Fig.~\ref{fig:density}b that the agreement between the electron density determinations from the MIR and from the optical lines improves as the SH/LH module scaling factor reaches large values ($\gtrsim0.4$). 
Note that the outlier N66\#11 is characterized by a highly uncertain density determination because of the extremely small [S\3] line ratio, in the low-density regime.

As a conclusion, the stitching of the SH and LH module spectra implies significant systematic uncertainties on the electron density determinations. We decided to use instead the values from the optical studies for the abundance determinations in \S\ref{sec:ionicab_results}. 
The electron temperature values were also taken from the optical analysis. 
We assumed $T_e=10\,000$\,K and $N_e=1\,000$\,cm$^{-3}$ for NGC\,3603 (Melnick et al.\ 1989; Garc{\'{\i}}a-Rojas et al.\ 2006), 
$T_e=10\,000$\,K and $N_e=100$\,cm$^{-3}$ for 30\,Dor (Kurt \& Dufour 1998; Peimbert 2003; 
Tsamis et al.\ 2003; Vermeij \& van der Hulst 2002), and $T_e=12\,500$\,K and $N_e=100$\,cm$^{-3}$ for N\,66 
(Kurt \& Dufour 1998; Peimbert et al.\ 2000). 
The electron density and temperature were assumed to be uniform across each object. The influence of possible variations of  $N_e$ and $T_e$ within the gaseous nebula is discussed in \S\ref{sec:ionicab_results}.

\subsection{Results and caveats}\label{sec:ionicab_results}

\begin{deluxetable}{llllllll}
\tabletypesize{\scriptsize}
\tablewidth{0pc}
\tablecolumns{8}
\tablecaption{Ionic abundances.\label{tab:ionicab}}
\startdata
\tableline\tableline
               & Ne\2/H    & Ne\3/H    & S\3/H     & S\4/H     & Ar\3/H  & Fe\2/H  & Fe\3/H   \\
\tableline
NGC3603\#3  & 5.44e-5 &	3.53e-5 &  4.86e-6 & 3.57e-7 & 2.60e-6 & 2.50e-7 & 3.37e-7      \\
NGC3603\#4  &  6.74e-5 & 1.89e-5 & 8.09e-6 & 1.94e-7 & 2.16e-6 & 4.06e-7 & 5.92e-7     \\ 
NGC3603\#5  & 4.27e-5 &	4.43e-5	 & 8.39e-6 & 6.53e-7 & 3.43e-6 & 3.53e-7 & 4.05e-7      \\
NGC3603\#6  & $<$4.8e-6 &7.16e-6 & $<$1.2e-6 & $<$1.4e-7 & $<$6.7e-7 & $<$1.1e-7  &$<$9.0e-8      \\
NGC3603\#7  & $<$1.4e-5 &$<$7.54e-5 & $<$7.1e-6 & $<$1.8e-6  &$<$2.3e-5 & $<$8.9e-7 & $<$2.0e-6      \\
NGC3603\#8  & $<$2.9e-5 &$<$2.00e-4 & \nodata\tablenotemark{a}  & $<$1.9e-6  &$<$5.6e-5 & $<$6.1e-6 & $<$1.6e-5      \\
NGC3603\#9  & $<$5.3e-5 & $<$1.82e-4 & $<$1.1e-5 & $<$1.8e-6 &$<$4.9e-5 & $<$4.e-10 & $<$1.1e-5      \\
\tableline
NGC\,3603 (average)  &   5.48e-5   & 2.64e-5    & 7.11e-6     & 4.01e-7  & 2.73e-6   & 3.36e-7   & 4.45e-7   \\
(dispersion, \%)    &  10.6 &  27.2         &    13.0       & 27.3          &  11.1      &    11.1       & 14.0     \\
\tableline
30DOR\#2   & 1.91e-5& 3.58e-5 & 5.32e-6& 4.06e-7& 1.75e-6 & 7.86e-8&  2.77e-7     \\ 
30DOR\#3   & 1.15e-5& 4.12e-5 & 4.88e-6& 8.06e-7& 1.51e-6&  4.83e-8&  2.09e-7      \\
30DOR\#4   & 7.24e-6& 4.72e-5&  4.43e-6& 1.48e-6& 1.57e-6 & $<$8.6e-8&  2.90e-7      \\
30DOR\#5   & 2.31e-5& 3.89e-5&  6.07e-6& 5.21e-7& 2.22e-6&  2.69e-7&  3.68e-7      \\
30DOR\#6   & 2.03e-5& 4.56e-5 & 6.48e-6& 7.00e-7& 1.64e-6&  4.62e-7&  5.86e-7     \\ 
30DOR\#7   & 9.11e-6&4.16e-5&  4.35e-6& 9.98e-7& 1.76e-6 & 1.04e-7&  2.81e-7      \\
30DOR\#8   & 2.84e-5&2.44e-5& 4.77e-6& 2.35e-7& 1.90e-6&  2.98e-7 & 4.05e-7      \\
30DOR\#10   & 4.05e-5& 1.55e-5&  5.01e-6& 2.17e-7& 1.72e-6&  2.93e-7&  3.66e-7     \\ 
30DOR\#11   & 1.79e-5& 4.46e-5 & 6.15e-6& 6.27e-7& 2.27e-6 & 1.96e-7&  4.03e-7     \\ 
30DOR\#12   & 1.35e-5& 4.83e-5&  5.47e-6& 1.07e-6& 2.38e-6&  2.43e-7&  3.58e-7      \\
30DOR\#13   & 1.38e-5& 3.75e-5&  5.11e-6& 6.27e-7& 2.07e-6 & $<$1.5e-7&  3.73e-7      \\
30DOR\#14   & 1.29e-5& 4.60e-5&  4.92e-6& 9.48e-7& 1.64e-6&  2.39e-7&  2.87e-7      \\
30DOR\#15   &  1.30e-5& 5.37e-5&	 5.88e-6& 1.08e-6& 2.34e-6 & 2.06e-7 & 3.46e-7    \\  
30DOR\#16   & 7.34e-6& 4.51e-5&  4.54e-6& 1.21e-6 &1.69e-6 & $<$1.4e-7&  2.45e-7     \\ 
30DOR\#17   & 3.53e-5& 1.54e-5&  4.18e-6& 1.23e-7& 2.70e-6&  1.40e-6&  1.18e-6      \\
\tableline
30\,Dor (average)  & 1.82e-5  &  3.87e-5 & 5.11e-6 & 7.37e-7    & 1.94e-6 &  3.20e-7 & 3.99e-7 \\
(dispersion \%)&  13.6       &   7.5         &   3.3        &  13.4       &   4.6       &   30.9      &  14.6   \\
\tableline
N66\#1   &  4.09e-6& 2.05e-5 & 2.51e-6& 6.31e-7& 7.12e-7&  $<$1.3e-7&  3.13e-8      \\
N66\#2   &  4.64e-6& 1.91e-5 & 2.30e-6& 5.99e-7& 5.37e-7&  $<$1.6e-7 & $<$1.2e-7      \\
N66\#3   & $<$2.0e-5  &  $<$6.72e-5  & $<$7.71e-6 &  $<$2.18e-6 & $<$4.12e-6&  $<$7.3e-7 & $<$6.8e-7      \\
N66\#5   &  3.52e-6&	1.44e-5&  1.72e-6& 3.65e-7& 1.18e-6&  $<$1.3e-7&  $<$2.0e-7     \\ 
N66\#6   &  8.69e-6&	1.60e-5& 2.26e-6& 4.99e-7& 2.06e-6&  5.80e-7 & $<$4.2e-7     \\ 
N66\#7   &  2.52e-6&	1.59e-5& 2.30e-6& 5.01e-7& 1.25e-6&  4.99e-7 & $<$3.9e-7     \\ 
N66\#8   &  6.92e-6&	1.69e-5	& 2.22e-6& 3.81e-7& 8.50e-7&  1.06e-7&  7.48e-8    \\  
N66\#9   &  7.69e-6&	1.70e-5	& 2.25e-6& 3.26e-7& 1.06e-6&  $<$1.9e-7&  $<$2.7e-7    \\  
N66\#10  &  6.92e-6&	1.52e-5& 2.00e-6 &2.96e-7& 9.52e-7&  $<$1.7e-7 & $<$2.4e-7     \\ 
N66\#11  &  8.05e-6&	1.76e-5& 2.42e-6 &3.41e-7& 1.11e-6&  $<$2.0e-7 & $<$2.8e-7     \\ 
N66\#12  &  2.59e-6&	1.38e-5	& 1.30e-6 &5.13e-7& 9.00e-7&  $<$8.9e-8&  $<$8.8e-8    \\  
N66\#13  &  3.87e-6&	1.40e-5	& 2.57e-6& 3.50e-7& 8.80e-7&  $<$1.4e-7&  $<$1.9e-7     \\ 
\tableline
N\,66 (average)  &          5.41e-6   &  1.64e-5  & 1.68e-6  & 4.36e-7 &  1.04e-6 & 3.95e-7 & 5.30e-7 \\
(dispersion \%)&   12.1      &  3.7         &   5.4       &  7.6        &    10.9    &   30.2    & 29.0 \\
\tableline
\enddata
\tablecomments{The measurement error on the ionic abundance is $\sim$15\% for Ne\2, Ne\3, S\3, S\4, Fe\2, and $\sim$20\% for Ar\3\ and Fe\3\ (\S\ref{sec:measurements}). The systematic error related to electron temperature variation is $\lesssim$20\% (\S\ref{sec:ionicab_results}).}
\tablenotetext{a}{Saturated [S\3] line.}
\end{deluxetable}

\begin{deluxetable}{lllllll}
\tabletypesize{\scriptsize}
\tablewidth{0pc}
\tablecolumns{4}
\tablecaption{Average ionic abundances compared to optical studies.\label{tab:ionicab_global}}
\startdata
\tableline\tableline
 Ne\2/H    & Ne\3/H    & S\3/H     & S\4/H     & Ar\3/H  & Reference & Comment  \\
\tableline
\multicolumn{7}{c}{NGC\,3603}   \\
 \tableline
5.48e-5 & 2.64e-5 & 7.11e-6 & 4.01e-7 & 2.73e-6 & (this study) & Average (3-4 positions) \\
  \nodata & 1.00e-4 & 1.29e-5 & \nodata  & 3.63e-6 & 1& Optical \\
\tableline
\multicolumn{7}{c}{30\,Dor}      \\
 \tableline
1.82e-5 & 3.87e-5 & 5.11e-6 & 7.37e-7 & 1.94e-6 & (this study) & Average (15 positions)\\
 \nodata& 5.89e-5 & 5.50e-6 &\nodata  & 1.51e-6 & 4 & Optical \\
 3.71e-5 &  \nodata      &4.28e-6 &\nodata &  1.11e-6 & 5 & Optical  \\
 2.33e-5 & 5.34e-5 &    7.56e-6  &    1.32e-6 & 1.34e-6& 2 & position \#1, ISO\\
  \nodata & \nodata &  3.52e-6  &    \nodata &  9.37e-7  & 2 & position \#1, Optical\\
 2.02e-5 & 6.04e-5 &     6.15e-6 & 1.46e-6 &1.14e-6&2 & position \#2, ISO\\
 \nodata & 3.63e-5 &   3.19e-6 &    \nodata &    8.85e-7 & 2 & position \#2, Optical\\
 2.03e-5 &  5.84e-5 &5.54e-6   &  1.48e-6 &1.45e-6&2 & position \#3, ISO\\
 \nodata  & 3.79e-5 &    4.36e-6   &    \nodata &  1.16e-6 & 2 & position \#3, Optical\\
 3.02e-5 & 7.37e-5 &8.73e-6    & 1.27e-6   &1.50e-6& 2 & position \#4, ISO\\
  \nodata & 3.88e-5 &    3.36e-6  &    \nodata &    8.46e-7  & 2 & position \#4, Optical\\
\tableline
\multicolumn{7}{c}{N\,66}   \\
\tableline
5.41e-6   & 1.64e-5 & 1.68e-6 & 4.36e-7 & 1.04e-6 & (this study) & Average (11 positions) \\
 1.27e-5 & 1.74e-5  & 1.30e-6 & 4.19e-7 & 6.57e-7 & 2 & N\,66, ISO\\
  \nodata   &    1.57e-5  &  1.53e-6      & \nodata & 3.81e-7 & 2 & N\,66, optical\\
  \nodata  & 1.35e-5   & 1.41e-6 &   \nodata & 3.89e-7& 3 & Region A, optical \\
\enddata
\tablerefs{(1) Garc{\'{\i}}a-Rojas et al.\ 2006; (2) Vermeij \& van der Hulst 2002; (3) Peimbert et al.\ 2000; (4) Peimbert 2003; (5) Tsamis et al.\ 2003.}
\end{deluxetable}

Ionic abundances assuming uniform electron density and temperature are presented in Table~\ref{tab:ionicab}. 
For comparison, we report in Table~\ref{tab:ionicab_global} the values derived from the optical.
Given the line flux uncertainties discussed in \S\ref{sec:measurements}, we consider the measurement error on the ionic abundance determination
to range from $\sim$15\% (for Ne\2, Ne\3, S\3, S\4, and Fe\2) to $\sim$25\% (for Ar\3\ and Fe\3).

\begin{figure*}
\figcaption{The ionic abundance determination is plotted as a function of $N_e$ and $T_e$. 
Densities are 100\,cm$^{-3}$ (circles), 1000\,cm$^{-3}$ (squares), and 10\,000\,cm$^{-3}$ (diamonds).
Results were normalized to the determinations assuming 100\,cm$^{-3}$ and $T_e=10\,000$\,K.
\label{fig:ionicab_var}}
\includegraphics[angle=0,scale=0.7,clip=true]{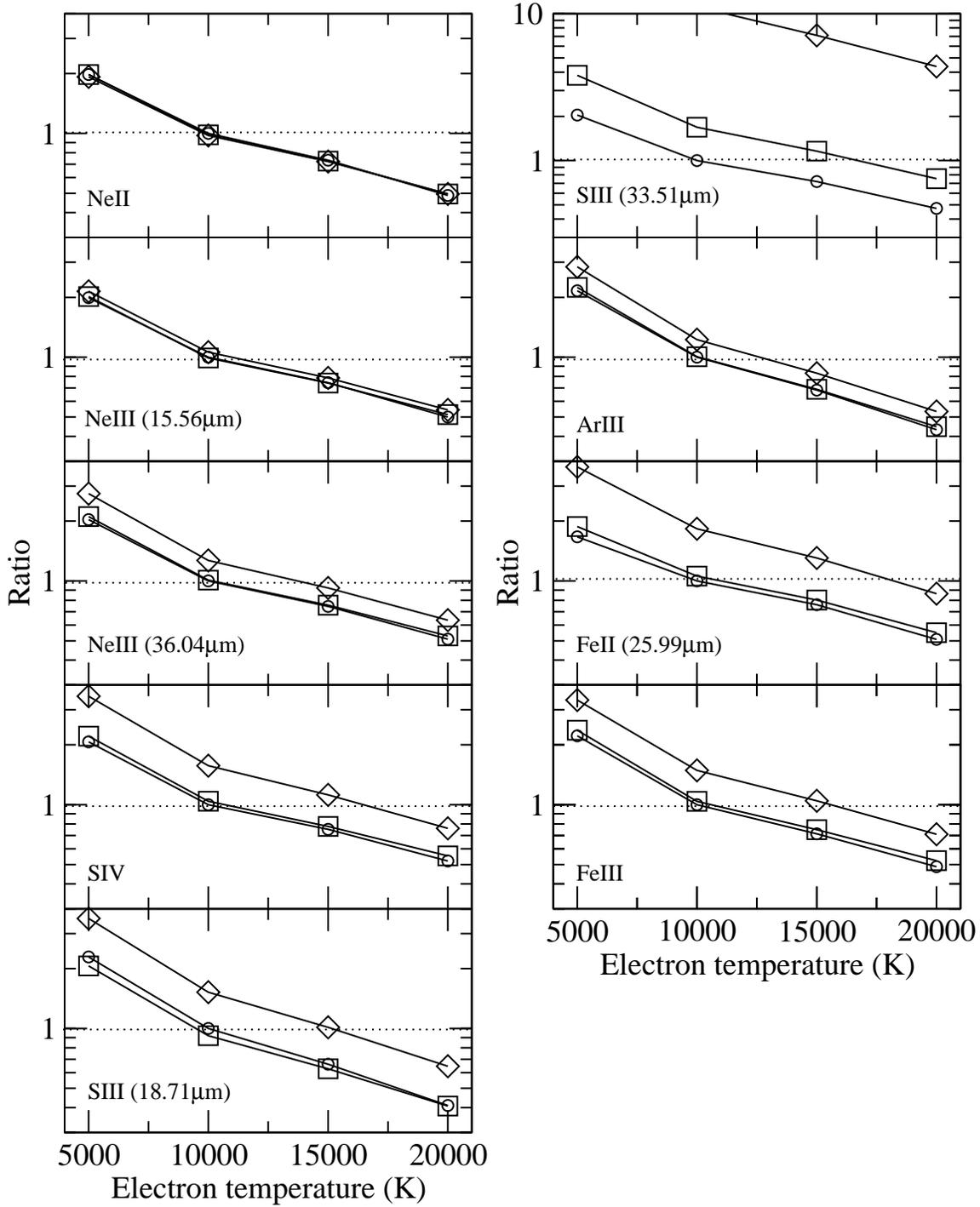}
\end{figure*}

Additional systematic errors on the method are due to the assumed physical conditions. It is likely that
$N_e$ and $T_e$ vary across a single giant H\2\ region, and along a given line of sight. 
It has been argued that $N_e$ in Galactic and extragalactic H\2\ regions is higher toward the brightest regions, reaching a few 100\,cm$^{-3}$, while the faintest Galactic H\2\ regions are characterized by uniform electron density on the order of 20-140\,cm$^{-3}$  (e.g., Castaneda et al.\ 1992; Copetti et al.\ 2000). 
It can be seen in Fig.\,\ref{fig:ionicab_var} that the ionic abundance determinations of Ne\2\ and Ne\3\ in the giant H\2\ regions of our sample are fairly insensitive to electron density in the range $10^2$-$10^4$\,cm$^{-3}$. The other ions do not show strong variations in their abundance for densities smaller than $10^3$\,cm$^{-3}$. 
Furthermore, low-excitation faint regions in N\,66 (N66\#1 and N66\#2) do not show ionic abundance
determinations particularly lower than the other $-$ brighter $-$ positions in this object. We conclude that internal variations (or uncertainties) in $N_e$ do not affect significantly the ionic abundance determinations.

On the other hand, variations of the electron temperature may have a significant impact on the abundance determinations. 
The uncertainty using different electron temperatures is much reduced when analyzing MIR fine-structure lines as compared to the use of lines in the optical spectrum (Bernard-Salas et al.\ 2001). 
In first approximation, the MIR ionic abundance determinations vary linearly with $T_e$. All the ionic abundances show similar trends with $T_e$ in Fig.\,\ref{fig:ionicab_var} because of the dependence of the recombination coefficient of H$\beta$ with $T_e$ (\S\ref{sec:ionicab_method}). 
In gaseous nebulae of giant H\2\ regions, the temperature is usually between 7\,500\,K and 15\,000\,K, while temperature fluctuations 
are on the order of 15-20\% (Esteban et al.\ 2002; O'Dell et al.\ 2003). We conclude that there might be a systematic error on the
ionic abundance determinations up to 20\% if $T_e$ is not uniform and vary by as much as 20\%.

The final uncertainty on the ionic abundances is given by the sum of measurement errors and errors due to electron temperature variation. 
Total errors range from $\pm$15\%$\pm$20\% ($\pm$0.06$\pm$0.08\,dex) for Ne\3, Ne\3, S\3, S\4, and Fe\2, to $\pm$20\%$\pm$20\% ($\pm$0.08$\pm$0.08\,dex) for Ar\3\ and Fe\3.
The dispersion of the ionic abundances we derive across the giant H\2\ regions ranges from $\sim$3\% to $\sim$30\%  (Table~\ref{tab:ionicab}), which is smaller than the total uncertainty ($\sim$35-40\%). This implies that (1) the errors on the ionic abundance could be somewhat overestimated, and (2) electron temperature are unlikely to vary by more than 20\% across a given giant H\2\ region.

\section{Elemental abundance determination}\label{sec:eleab}

The MIR range gives the unique opportunity to observe the most important ionization stages of elements such as Ne, S, and Ar.
As a result, the elemental abundance determination requires no $-$ or little $-$ ionization correction factors. 
Iron abundance determination is much more uncertain because we do not observe the dominant ion in the ionized gas (Fe\4). 
Finally, we could not measure the total abundance of silicon, since we have access only to the Si\2\ ion.

\begin{deluxetable}{lrrrrr}
\tabletypesize{\scriptsize}
\tablecolumns{6}
\tablewidth{0pc}
\tablecaption{Ionization potentials.\label{tab:ip}}
\startdata
\tableline
\tableline
    & \1 &\2 &\3 & \4 & \5 \nl
\tableline
H  & 13.60 &   \nodata     &    \nodata     &   \nodata     &    \nodata   \nl
Ne & 21.56 & 40.96  & 63.45  &  97.11& 126.21 \nl
Si & 8.15  &16.34   &  33.49 & 45.15 & 166.77 \nl
S  & 10.36 &23.33   &  34.83 &47.30  & 72.68  \nl
Ar & 15.76 &27.63   & 40.74  & 59.81 &  75.02 \nl
Fe & 7.87  & 16.18  & 30.64  & 54.8  & 75.0   \nl
\enddata
\tablecomments{The ionization potentials are expressed in eV.}
\end{deluxetable}

\subsection{Sulfur ionization structure}\label{sec:sulfurdet}

The total abundance of sulfur was calculated using the sum of the S\3\ and S\4\ ionization stages.
We used the [S\3] line at 18.71\mic\ instead of the one at 33.48\mic\ to estimate the ionic abundance of S\3, because the 18.71\mic\ line is measured in the same module as H\1~$_{\rm 7-6}$, hence allowing us to avoid aperture effects. In addition, the 18.71\mic\ line is much less
sensitive to electron density variations (Fig.~\ref{fig:ionicab_var}).
No ionization corrections were made due to the presence of other ionization stages.
We cannot exclude that some S\2\ is present in the ionized gas given the ionization potential (IP) of S\2\ (Table~\ref{tab:ip}). 
The photoionization cross-section (PICS) of S\1\ is actually higher than that of S\2\ for energies $\lesssim$50\,eV (Verner et al.\ 1996). 
It is however unknown which fraction of S\2\ resides in the ionized gas, in the neutral gas, or in the associated PDRs.

Photoionization models as well as optical observations support the predominancy of S\3\ and S\4\ stages.
Models of Tsamis et al.\ (2005) for 30\,Dor predict
that $\sim$92\% of sulfur is into S\3\ and S\4, while 8\% is due to other ionization stages, mostly S\2.
Peimbert (2003) and Vermeij \& van der Hulst (2002) find consistent results observationaly, with S\2/S\3\ ranging from 3\% to 8.5\%.
In N\,66, the contribution of S\2\ is approximately 15\%\ that of S\3\ (Peimbert et al.\ 2000; Vermeij \& van der Hulst 2002). 
Finally, in NGC\,3603, S\2/S\3 is only 1.2\%\ (Garc{\'{\i}}a-Rojas et al.\ 2006). 
Hence, we might underestimate the sulfur abundance by $\sim$10\%\ in 30\,Dor and N\,66 while the ionization correction factor (ICF) in NGC\,3603 is negligible.

The final uncertainty on S/H is due to measurement errors on the ionic abundances of S\3\ and S\4\ (15\% each; \S\ref{sec:measurements}), and to the assumed physical conditions ($\lesssim$20\%, which affect S\3\ and S\4\ the same way, \S\ref{sec:ionicab_results}). 

\subsection{Neon ionization structure}\label{sec:neondet}

The abundance of neon was calculated using both Ne\2\ and Ne\3\ ions. We use the [Ne\3] line at 15.56\mic\ instead of the one at 36.01\mic\ because it resides in the same module as the H\1~$_{\rm 7-6}$ line, and because the 15.56\mic\ line is comparatively less sensitive to electron density variations (Fig.~\ref{fig:ionicab_var}).
As far as higher ionization stages are concerned, the contribution of Ne\4\ is expected to be negligible since the IP of Ne\3\ is 63.45\,eV, i.e., above the He\2\ absorption edge at 54.4\,eV in ionizing stars.
The PICS of Ne\3\ is actually smaller than that of Ne\2\ for energies $\lesssim$70\,eV (Verner et al.\ 1996), implying that Ne\4\ should not exist in significant amount. 
Moreover, the [O\4] line at 25.89\mic\ is not detected in any of our spectra but one which will be discussed in Lebouteiller
et al.\ (in preparation). 
The IP of O\3\ is 54.9\,eV, as compared to 63.4\,eV for Ne\3, hence the absence of O\4\ implies the absence of Ne\4.

Models of 30\,Dor confirm these findings and predict that Ne\4\ represents $\sim0.0002$\% of the total neon, as compared to 86\%\ for Ne\3\ and 14\%\ for Ne\2\ (Tsamis et al.\ 2005). 
The proportion of Ne\4\ should be negligible also in N\,66 and NGC\,3603, resulting in similar negligible ionization corrections.
More specifically, given the fact that the global interstellar radiation field (ISRF) hardness in N\,66 and NGC\,3603 is similar or lower than that in 30\,Dor (Lebouteiller et al.\ in preparation), we do not expect any significant contribution of Ne\4\ in any of our objects.

Similarly to sulfur abundance (\S\ref{sec:sulfurdet}), the error on Ne/H is $\pm$30\% due to measurement errors, and $\lesssim$20\% due to the assumed physical conditions.

\subsection{Argon ionization structure}\label{sec:argondet}

\begin{deluxetable}{llll}
\tabletypesize{\scriptsize}
\tablewidth{0pc}
\tablecolumns{4}
\tablecaption{[Ar\2] line flux and Ar\2\ ionic abundance.\label{tab:ionicab_arii}}
\startdata
\tableline\tableline
               & [Ar\2]\tablenotemark{a}    & Ar\2/H\tablenotemark{b}    & Ar\2/Ar\3\    \\
\tableline
NGC3603\#3  & 6.90 & 1.79e-7 & 6.9\%  \\
NGC3603\#4  & 26.20 & 4.05e-7 & 18.8\%  \\ 
NGC3603\#5  & 4.37  & 1.34e-7 & 3.9\%   \\
NGC3603\#6  & 2.96 & 7.69e-8 & \nodata\tablenotemark{c}    \\
NGC3603\#7  &  \nodata& \nodata&   \nodata\tablenotemark{c}    \\
NGC3603\#8  & 4.47 & 1.91e-8 & \nodata\tablenotemark{c}    \\
NGC3603\#9  &  \nodata& \nodata&  \nodata\tablenotemark{c}   \\
\tableline
NGC\,3603 (average)   & \nodata & 1.63e-7 & 9.9\%   \\
\tableline
30DOR\#2   & 0.11 & 4.24e-8 & 2.4\%   \\ 
30DOR\#3   & 1.56 & 1.44e-7 & 9.5\%   \\
30DOR\#4   &  \nodata &  \nodata &   \nodata   \\
30DOR\#5   & 0.12 & 2.99e-8 & 1.3\%   \\
30DOR\#6   & 0.33 & 6.08e-8 & 3.7\%    \\ 
30DOR\#7   & 0.91 & 3.54e-8 &  2.0\%   \\
30DOR\#8   & 2.45 & 1.30e-7 & 6.8\%  \\
30DOR\#10   & 2.34 & 1.54e-7  & 9.0\%  \\ 
30DOR\#11   & 0.10 & 7.99e-9 &  0.3\%  \\ 
30DOR\#12   &  \nodata & \nodata &  \nodata    \\
30DOR\#13   &  \nodata&  \nodata &   \nodata   \\
30DOR\#14   & 1.54 & 4.59e-8 &  2.8\%  \\
30DOR\#15   &  0.66 & 1.94e-8 & 0.8\%  \\  
30DOR\#16   & 0.30 & 9.93e-9 & 0.6\%   \\ 
30DOR\#17   & 0.24 & 7.10e-8 &  2.7\%   \\
\tableline
30\,Doradus (average)   & \nodata & 6.26e-8 & 3.5\%    \\
\tableline
N66\#1   &   \nodata &  \nodata & \nodata     \\
N66\#2   &  \nodata &  \nodata &  \nodata    \\
N66\#3   &  \nodata &  \nodata &   \nodata    \\
N66\#5   &  0.06 & 3.23e-8 &  2.7\%   \\  
N66\#6   &  0.10 & 7.66e-8 & 3.7\%    \\ 
N66\#7   &  \nodata&  \nodata&  \nodata   \\ 
N66\#8   &  0.03 & 1.71e-8 &  2.0\%  \\  
N66\#9   &  0.12 & 1.05e-7 &  9.9\%  \\  
N66\#10  &  0.09 & 6.95e-8 &  7.3\%   \\ 
N66\#11  &  \nodata&  \nodata &   \nodata \\ 
N66\#12  &  0.03 & 1.61e-8 & 1.8\%   \\  
N66\#13  &  0.13 & 8.21e-8 &  9.3\%  \\ 
\tableline
N\,66 (average)   & \nodata & 6.55e-8 & 5.2\%    \\
\tableline
\enddata
\tablenotetext{a}{[Ar\2] line flux is in units of $\times10^{-20}$\,W\,cm$^{-2}$. We estimate the measurement error to be $\sim20$\%.}
\tablenotetext{b}{The measurement error on the ionic abundance is $\sim$25\% (\S\ref{sec:argondet}). The systematic error related to electron temperature variation is $\lesssim$20\% (\S\ref{sec:ionicab_results}).}
\tablenotetext{c}{Only an upper limit could be determined for the Ar\3\ ionic abundance (Table~\ref{tab:ionicab}).}
\end{deluxetable}

Given their IP (Table~\ref{tab:ip}), we expect \textit{a priori} Ar\2, Ar\3, and Ar\4\ to be present in the ionized gas.
Through the SL module of the IRS, we have access to the [Ar\2] line at 6.99\mic. We were able to estimate the Ar\2\ ionic abundance using the method described in \S\ref{sec:ionicab_method}, with an estimated uncertainty as large as 25\%, principally due to the low spectral resolution of the SL module together with the faintness of the line.
The results show that the Ar\2/Ar\3\ abundance ratio is smaller than 10\% except toward the position \#4 in NGC\,3603 (Table~\ref{tab:ionicab_arii}). The line of sight toward NGC3603\#4 is characterized by a relatively low ionization degree, based on the (Ne\3/H)/(S\3/H) ionic abundance ratio (\S\ref{sec:iondegree}), which explains the large amount of Ar\2. 
It is not entirely clear, however, whether Ar\2\ belongs to the ionized gaseous phase. In fact, even though Ar\1\ has an IP of 15.76\,eV, i.e., above 
that of H\1, Sofia \& Jenkins (1998) proposed that the PICS of Ar\1\ is such that it can be ionized into Ar\2\ 
when hydrogen is still in H\1. The calculations of chemical abundances in the neutral gas of star-forming regions seem to agree with 
this proposal (see e.g., Lecavelier et al.\ 2004; Lebouteiller et al.\ 2005, 2006). In addition, the study of the spatial variations of MIR features across NGC\,3603 shows that the [Ar\2] line intensity correlates with that of the PDR tracers such as the PAHs, and not with usual ionized gas tracers such as [Ne\2], [Ne\3], [S\3], or [S\4] (Lebouteiller et al.\ 2007).
Hence, the contribution of Ar\2\ we derived should be regarded as an upper limit.

\begin{figure}[h!]
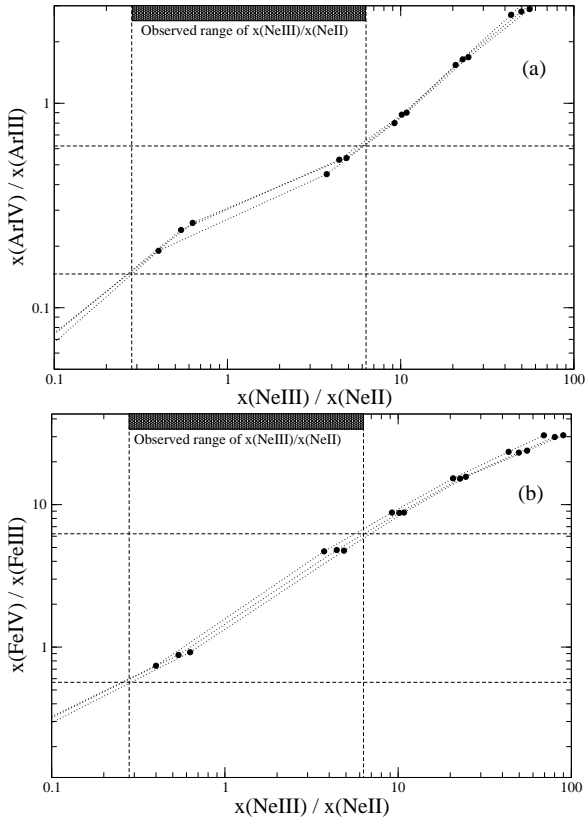

\includegraphics[angle=0,scale=0.32,clip=true]{f8a.eps}\\
\includegraphics[angle=0,scale=0.32,clip=true]{f8b.eps}\\
\figcaption{Ionic fractions based on the model grids of Stasinska et al.\ (1990). Models considered were (abcdefg)2(bcd)1.
\label{fig:ioncorr}}
\end{figure}

The IP of Ar\3\ (40.74\,eV) is similar to that of Ne\2\ (40.96\,eV), thus it could be expected that, together with Ar\2\ and Ar\3, Ar\4\ is present in the ionized gas, given the ubiquitous detection of [Ne\3] in our spectra. 
We consider however that Ar\4\ contribution is not significant given the relatively low PICS of Ar\3\ as compared to that of Ar\2\ for energies smaller than $\sim$70\,eV (Verner et al.\ 1996).
Optical observations of NGC\,3603 show that Ar\4\ contribution is indeed negligible, with Ar\4$\approx0.025\times$Ar\3\ (Garc{\'{\i}}a-Rojas et al.\ 2006). 
A similar value is found for 30\,Dor, where
the contribution of Ar\4\ is $\sim$2-3\%\ that of Ar\3\ (Peimbert 2003; Tsamis et al.\ 2003).
In N\,66 on the other hand, Ar\4\ could represent as much as 17\%\ as compared to Ar\3\ (Peimbert et al.\ 2000). Such a large contribution is not surprising given the relatively hard ISRF in this object.
To investigate in more detail the presence of Ar\4, we used the photoionization grids of Stasi{\'n}ska (1990). Figure~\ref{fig:ioncorr}a shows the correlation between the ionization fraction ratios $x$(Ar\4)/$x$(Ar\3) and $x$(Ne\3)/$x$(Ne\2). In the spectra of the three giant H\2\ regions of our sample, $x$(Ne\3)/$x$(Ne\2) ranges from $\approx$0.28 to $\approx$6.31 (dashed lines in Fig.~\ref{fig:ioncorr}a). 
This corresponds to $x$(Ar\4)/$x$(Ar\3) ranging from $\approx$0.15 to $\approx$0.64. 
Hence according to the models, Ar\3\ is the dominant stage, and corrections factors on the final argon abundance are on the order of 15-65\% (0.06-0.21\,dex). 
These ionization correction factors seem to be significantly larger than found in the optical (Ar\4/Ar\3$<$17\%, i.e., $<$0.07~dex). 
The optical spectra probe a gas with an ionization degree equal or higher as compared to the gas probed in the MIR (\S\ref{sec:iondegree}), hence the ICF due to the presence of Ar\4\ is likely smaller than $\sim$20\% in the MIR spectra.
The final results suggest that corrections are in fact negligible (\S\ref{sec:argondisc}).

The final error on Ar/H is due to measurement errors on the ionic abundances of Ar\2\ ($\sim$25\%) and Ar\3\ ($\sim$20\%), and to the assumed physical conditions ($\lesssim$20\%). 

\subsection{Iron ionization structure}\label{sec:irondet}

Given the IP of the iron ionization stages (see Table~\ref{tab:ip}), it is expected that most of the iron in the ionized gas is into Fe\2, Fe\3, and Fe\4. 
Fe\4\ is not expected to be further ionized because of the helium absorption edges in stars at 54.4\,eV.
We have access to both Fe\2\ (25.99\mic) and Fe\3\ (22.92\mic) in the MIR. 
The Fe\2\ ion does not necessarily arise in the ionized gas. In fact, we find that the [Fe\2] line flux correlates best with PDR tracers such as the PAH 
emission or [Ar\2] (Lebouteiller et al.\ in preparation). 

On the other hand, the presence of Fe\4\ could seriously hamper the iron abundance determination. 
One way to probe the presence of Fe\4\ is to study the variation of 
the ionic abundance ratio Fe\3/Fe\2 as a function of the ionization degree (see the application in M\,42 and M\,17 by Rodr{\'i}guez 2002). 
Fe\3/Fe\2\ is expected to decrease when the ionization becomes harder,
due to the presence of Fe\4. In our sample, there is no clear correlation with the ionization degree as probed by [Ne\3]/[Ne\2]. This means either that
Fe\4\ is not present in significant amount or that on the contrary it is the dominant stage over the whole range of physical conditions (density, ISRF hardness) in the H\2\ regions.

We estimated the ionic abundance of Fe\4\ using the photoionization grids of Stasi{\'n}ska (1990).
It can be seen in Fig.~\ref{fig:ioncorr}b that $x$(Fe\4)/$x$(Fe\3) lies between $\approx$0.56 and $\approx$6.21. 
Hence Fe\4\ represents a significant fraction of the total iron in the ionized gas, even becoming the dominant ionization stages in regions with [Ne\3]/[Ne\2] greater than $\approx$0.6, i.e., for most of our observations. In their optical study of NGC\,3603, Garc{\'{\i}}a-Rojas et al.\ (2006) assumed that Fe\4$\approx2.4\times$(Fe\2+Fe\3), confirming the predominancy of the Fe\4\ ion.
The correction we applied due to the presence of Fe\4\ ranges from a factor 0.6 to 6 (Table~\ref{tab:eleab}).

The error on Fe/H is due to measurement error on Fe\2/H (15\%) and Fe\3/H (20\%), and to the assumed physical conditions ($\lesssim$20\%). 
We corrected Fe/H for the presence of Fe\4. The error on the ICF is estimated to be as much as a factor 2.

\section{Discussion}\label{sec:discussion}

\begin{deluxetable}{lllll}
\tabletypesize{\scriptsize}
\tablewidth{0pc}
\tablecolumns{5}
\tablecaption{Elemental abundances from the MIR.\label{tab:eleab}}
\startdata
\tableline\tableline
               & Ne/H      & S/H      & Ar/H\tablenotemark{a}  &  Fe/H\tablenotemark{b}  (ICF)   \\
\tableline       
NGC3603\#3     & 7.95   & 7.08   & 6.41    & 5.98 (0.20)    \\
NGC3603\#4     & 7.94   & 6.92   & 6.33    & 6.13 (0.13)    \\
NGC3603\#5     & 7.94   & 6.96   & 6.53    & 6.14 (0.26)    \\
NGC3603\#6     & \nodata       & \nodata       & \nodata        &  \nodata     \\
NGC3603\#7     & \nodata       & \nodata       & \nodata        & \nodata       \\
NGC3603\#8     & \nodata       &  \nodata      & \nodata        & \nodata       \\
NGC3603\#9     & \nodata       & \nodata       & \nodata        & \nodata      \\
Error\tablenotemark{c} & $\pm0.11\pm0.08$  & $\pm0.11\pm0.08$ & $\pm0.16\pm0.08$   & $\pm0.37\pm0.08$ \\
\tableline
NGC\,3603 (average)   & 7.94 & 6.99 & 6.43  &  6.09 (0.19)         \\
(standard deviation)    & 0.00 &  0.09 &  0.11 &  0.09   \\
\tableline       
30DOR\#2       & 7.74   & 6.76  & 6.25    & 6.01 (0.46) \\
30DOR\#3       & 7.72   & 6.76  & 6.22    & 6.04 (0.63) \\
30DOR\#4       & 7.73   & 6.77  & 6.20    & 6.32 (0.86) \\
30DOR\#5       & 7.79   & 6.82  & 6.36    & 6.16 (0.36) \\
30DOR\#6       & 7.82   & 6.86  & 6.23    & 6.42 (0.40) \\
30DOR\#7       & 7.73   & 6.73  & 6.26    & 5.80 (0.21) \\
30DOR\#8       & 7.72   & 6.70  & 6.31    & 6.10 (0.25)  \\
30DOR\#10      & 7.75   & 6.72  & 6.28    & 5.98 (0.16) \\
30DOR\#11      & 7.80   & 6.83  & 6.36    & 6.26 (0.48) \\
30DOR\#12      & 7.79   & 6.82  & 6.38    & 6.32 (0.54) \\
30DOR\#13      & 7.71   & 6.76  & 6.32    & 6.20 (0.63) \\
30DOR\#14      & 7.77   & 6.77  & 6.22    & 6.24 (0.52) \\
30DOR\#15      & 7.82   & 6.84  & 6.37    & 6.33 (0.59)  \\
30DOR\#16      & 7.72   & 6.76  & 6.23    & 6.22 (0.83)  \\
30DOR\#17      & 7.71   & 6.63  & 6.44    & 6.57 (0.16) \\
Error\tablenotemark{c} & $\pm0.11\pm0.08$  & $\pm0.11\pm0.08$ &  $\pm0.16\pm0.08$   &  $\pm0.37\pm0.08$  \\
\tableline
30\,Dor (average)  & 7.76 & 6.77 & 6.32  & 6.24 (0.35)  \\
(standard deviation)    &  0.02  &  0.03 & 0.09 &  0.11    \\
\tableline         
N66\#1        & 7.39   & 6.50   & 5.85  & 5.30 (0.80) \\
N66\#2        & 7.37   & 6.46   & 5.73  & \nodata         \\
N66\#3        & \nodata&\nodata &\nodata& \nodata          \\
N66\#5        & 7.25   & 6.32   & 6.07  & \nodata        \\
N66\#6        & 7.39   & 6.44   & 6.31  & \nodata          \\
N66\#7        & 7.27   & 6.45   & 6.10  & \nodata         \\
N66\#8        & 7.38   & 6.42   & 5.93  & 5.61  (0.35) \\
N66\#9        & 7.39   & 6.41   & 6.03  & \nodata         \\
N66\#10       & 7.34   & 6.36   & 5.98  & \nodata       \\
N66\#11       & 7.41   & 6.44   & 6.05  & \nodata       \\
N66\#12       & 7.21   & 6.26   & 5.95  & \nodata       \\
N66\#13       & 7.25   & 6.33   & 5.94  & \nodata        \\
Error\tablenotemark{c}  & $\pm0.11\pm0.08$  & $\pm0.11\pm0.08$ &  $\pm0.16\pm0.08$   &  $\pm0.37\pm0.08$        \\
\tableline
N\,66 (average)        & 7.34 & 6.36   & 5.97  & 5.48 (0.45) \\
(standard deviation)& 0.05 & 0.10   &  0.15  & 0.24  \\
\enddata
\tablecomments{The abundance of an element X is expressed as 12+$\log$\,(X/H), where X/H is the sum of the ionic abundances.}
\tablenotetext{a}{Calculated using the ionic abundances of Ar\2\ and Ar\3.}
\tablenotetext{b}{Calculated using the ionic abundances of Fe\2, Fe\3, and Fe\4. The ICF applied due to the presence of Fe\4\ is indicated
between parentheses.}
\tablenotetext{c}{The first term gives the measurement error and the error on the ICF if applied. The second term gives the error
due to the assumed physial conditions (\S\ref{sec:ionicab_results}).}
\end{deluxetable}

 \begin{deluxetable}{lllllll}
 \tabletypesize{\scriptsize}
 \tablewidth{0pc}
 \tablecolumns{4}
 \tablecaption{Elemental abundances from the optical.\label{tab:eleab_opt}}
 \startdata
 \tableline\tableline
  Ne/H & S/H & Ar/H & Fe/H & Reference  \\
 \tableline
 \multicolumn{5}{c}{NGC\,3603}   \\
 \tableline
  $8.03\pm0.11$ & $7.36\pm0.08$ & $6.58\pm0.17$  &  $6.05\pm0.10$& (1) \\
  $7.94\pm0.26$ & $6.83\pm0.24$ & $6.55\pm0.13$  &  \nodata      & (2) \\
  $8.08$        & $7.12\pm0.09$ & \nodata        &  \nodata      & (3) \\
\multicolumn{3}{l}{Average Ne/S = 0.82$\pm$0.21} &    &  \\
 \tableline
 \multicolumn{5}{c}{30\,Dor}      \\
 \tableline
   \nodata              & $6.70\pm0.14$ & \nodata         & \nodata      & (3) \\
 $7.83\pm0.06$ & $6.99\pm0.10$ & $6.26\pm0.10$   & $6.39\pm0.20$& (4) \\
 $7.73$        & $6.84$        & $6.14$          & $6.12$       & (5) \\
 $7.88$ & $6.84$ & $6.15\pm0.07$ & \nodata & (6, pos\#1)\\
 $7.84\pm0.08$ & $6.80\pm0.09$ & $6.08\pm0.05$  & \nodata & (6, pos\#2) \\
 $7.84\pm0.16$ & $6.82\pm0.04$& $6.21\pm0.05$ & \nodata & (6, pos\#3) \\
 $7.94\pm0.07$ & $6.88\pm0.13$& $6.11\pm0.10$ & \nodata  & (6, pos\#4) \\
 $7.55$ & $6.71$& $6.03$ &\nodata & (7)\\
 $7.78\pm0.06$ & $6.67\pm0.01$ & $6.25$ & \nodata & (8) \\
\multicolumn{3}{l}{Average Ne/S = 1.03$\pm$0.12}    &  &  \\
 \tableline
 \multicolumn{5}{c}{N\,66}   \\
 \tableline
$7.47\pm0.01$ & $6.32\pm0.02$&  5.81  & \nodata   & (6) \\
$7.24$ &  \nodata         & \nodata          &  \nodata          & (9, N66A) \\
$7.20$ &  \nodata         &  \nodata         &  \nodata          & (9, N66NW) \\
$7.22$ &  \nodata         &  \nodata         &  \nodata          & (10) \\
$7.22$ & $6.50$     &    $5.74$     &  \nodata   & (12, t$^2$=0.0013) \\
\multicolumn{3}{l}{Average Ne/S = 1.15$\pm$0.02} &     &  \\
 \enddata
\tablecomments{The average Ne/S values were calculated from measurements with quoted errors only.}
 \tablerefs{(1) Garc{\'{\i}}a-Rojas et al.\ (2006); (2) Tapia et al.\ (2001); (3) Simpson et al.\ (1995); (4) Peimbert 2003; (5) Tsamis \& Pequignot 2005; (6) Vermeij \& van der Hulst (2002); (7) Rosa \& Mathis 1987; (8) Mathis et al.\ 1985; (10) Dufour, Shields \& Talbot (1982); (11) Dufour \& Harlow (1977); (12) Peimbert et al.\ (2000).}
 \end{deluxetable}

\begin{figure}[h!]
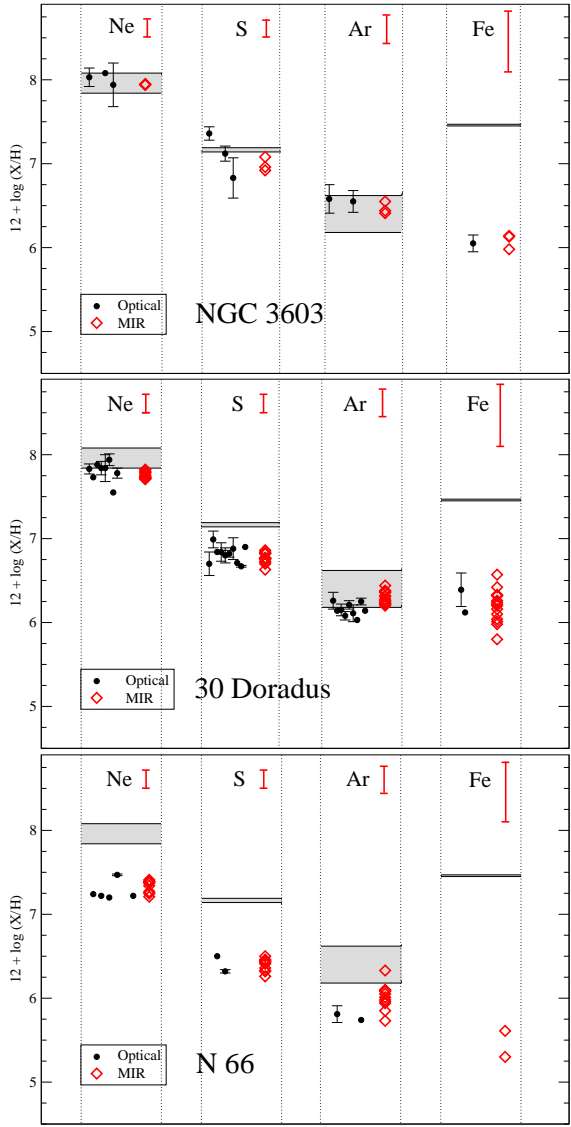

\includegraphics[angle=0,scale=0.325,clip=true]{f7a.eps}\\
\includegraphics[angle=0,scale=0.325,clip=true]{f7b.eps}\\
\includegraphics[angle=0,scale=0.325,clip=true]{f7c.eps}
\figcaption{Elemental abundances in NGC\,3603, 30\,Dor, and N\,66. Red diamonds correspond to MIR abundances while black dots refer to measurements in the optical (references are those given in \S\ref{sec:abopt}). The gray stripes indicate the range of solar abundance determinations for each element. Iron abundance is calculated using an ICF while argon abundance is calculated using only Ar\3\ ionic abundance (see \S\ref{sec:argondisc}). The error bars displayed close to the element label indicate the typical uncertainty on the corresponding abundance (0.16\,dex).
\label{fig:abundances}}
\end{figure}

Elemental abundances are presented in Table~\ref{tab:eleab} and illustrated in Fig.~\ref{fig:abundances}. 
The uncertainties are comparable to the abundance dispersion across each giant H\2\ region (Table~\ref{tab:eleab}), suggesting that the abundance 
variations we observe are mostly driven by uncertainties and are probably not intrinsic (see also discussion in \S\ref{sec:dispersion}).

Elemental abundances from optical studies are summarized in Table~\ref{tab:eleab_opt}. 
The MIR abundances of neon, sulfur, argon, and iron scale according to their abundance in the Sun.
For the three objects in our sample, the elemental abundances derived from MIR collisionally-excited lines (CELs; red diamonds in Fig.~\ref{fig:abundances}) agree globally with those derived from optical CELs (black dots). Because our abundance measurements are less affected by electronic temperature determinations, this suggests that the discrepancy often observed between optical recombination lines and CELs is not principally due to temperature fluctuations (see also Liu et al.\ 2001; Bernard-Salas et al.\ 2001).

\subsection{MIR abundances}

\subsubsection{Neon}\label{sec:neondisc}

The abundance of neon does not suffer from any significant uncertainty due to ionization corrections (\S\ref{sec:neondet}). 
Neon stands out as being the element showing the best agreement with the corresponding optical references (\S\ref{tab:eleab_opt}). 
The average abundance we derive in NGC\,3603 ($\approx7.94\pm0.11\pm0.08$) is relatively close to the value Ne/H=8.08 found by Simpson et al.\ (1995) using the same MIR CELs as we observe, but combining different observations with possible aperture effects. 

Neon is also the element which shows the smallest scatter in its abundance across the various positions in each giant H\2\ region.
As an illustration, the 15 positions in 30\,Doradus show only 0.11\,dex dispersion, the 3 positions in NGC\,3603 give remarkably equal values, while the 11 positions in N\,66 show 0.20\,dex dispersion.
These findings, together with the fact that neon is not expected to be depleted on dust grains or incorporated in molecules makes
it a reliable metallicity tracer, contrary to sulfur, argon, and iron (see also \S\ref{sec:cosmic}). For this reason, we consider it as a reference for our discussion.

\subsubsection{Sulfur}\label{sec:sulfurdisc}

The average abundances of sulfur we determined in the giant H\2\ regions corroborate the optical values (Fig.~\ref{fig:abundances}).  
There might be an indication however that our determinations agree best with the lowest optically-derived abundances (see discussion in 
\S\ref{sec:diffdep}).

The sulfur abundance shows a larger dispersion across each giant H\2\ region than neon.
Part of the larger dispersion could be attributed \textit{a priori} to ionization corrections due to the presence of S\2. However, an
ICF would change S/H by less than 0.05\,dex (assuming S\2/S\3=0.1; \S\ref{sec:sulfurdet}), which is smaller than the total uncertainty on the sulfur abundance determination ($\pm0.11\pm0.08$\,dex). 

\begin{figure}[h!]
\includegraphics[angle=0,scale=0.45,clip=true]{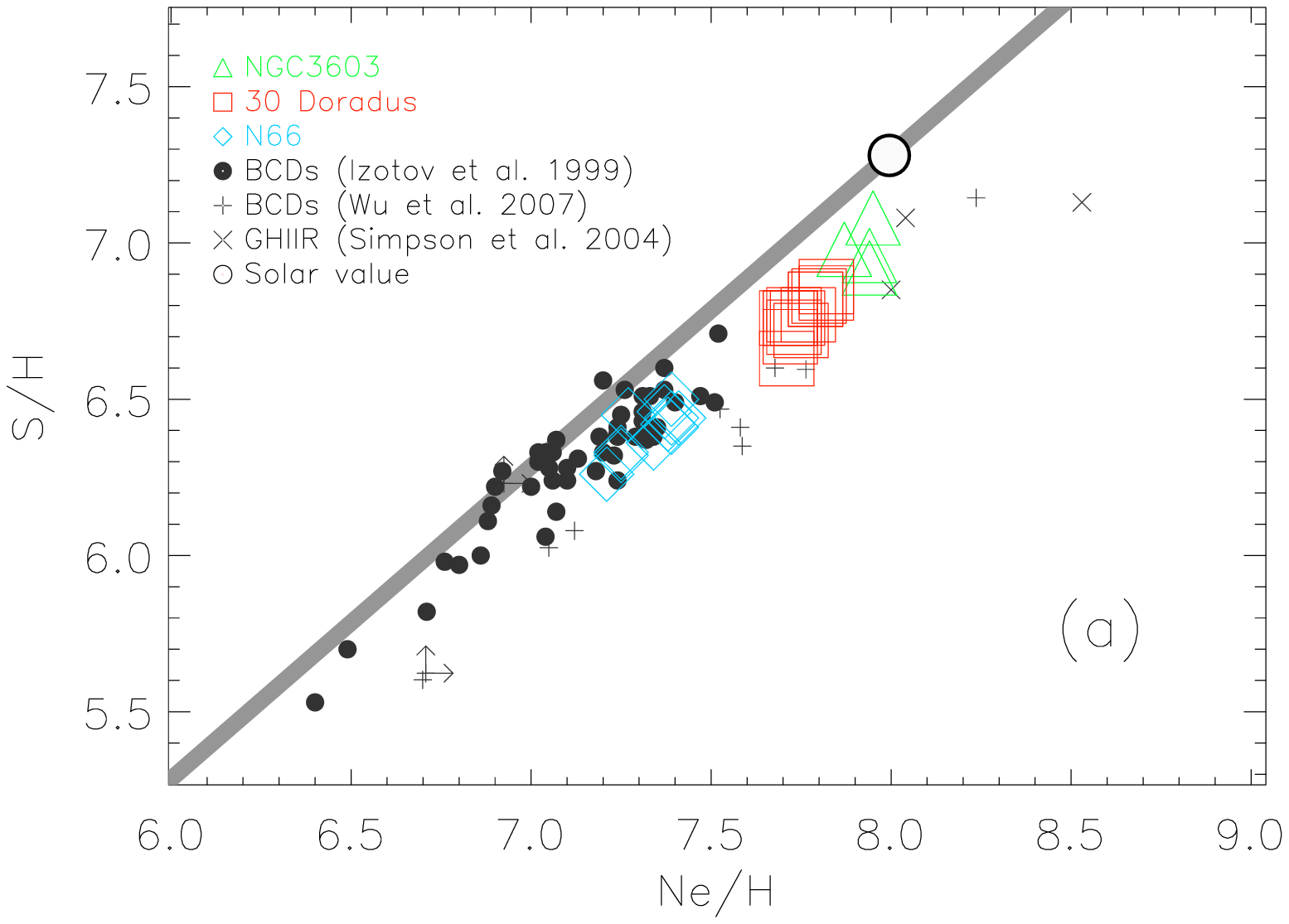}\\
\includegraphics[angle=0,scale=0.45,clip=true]{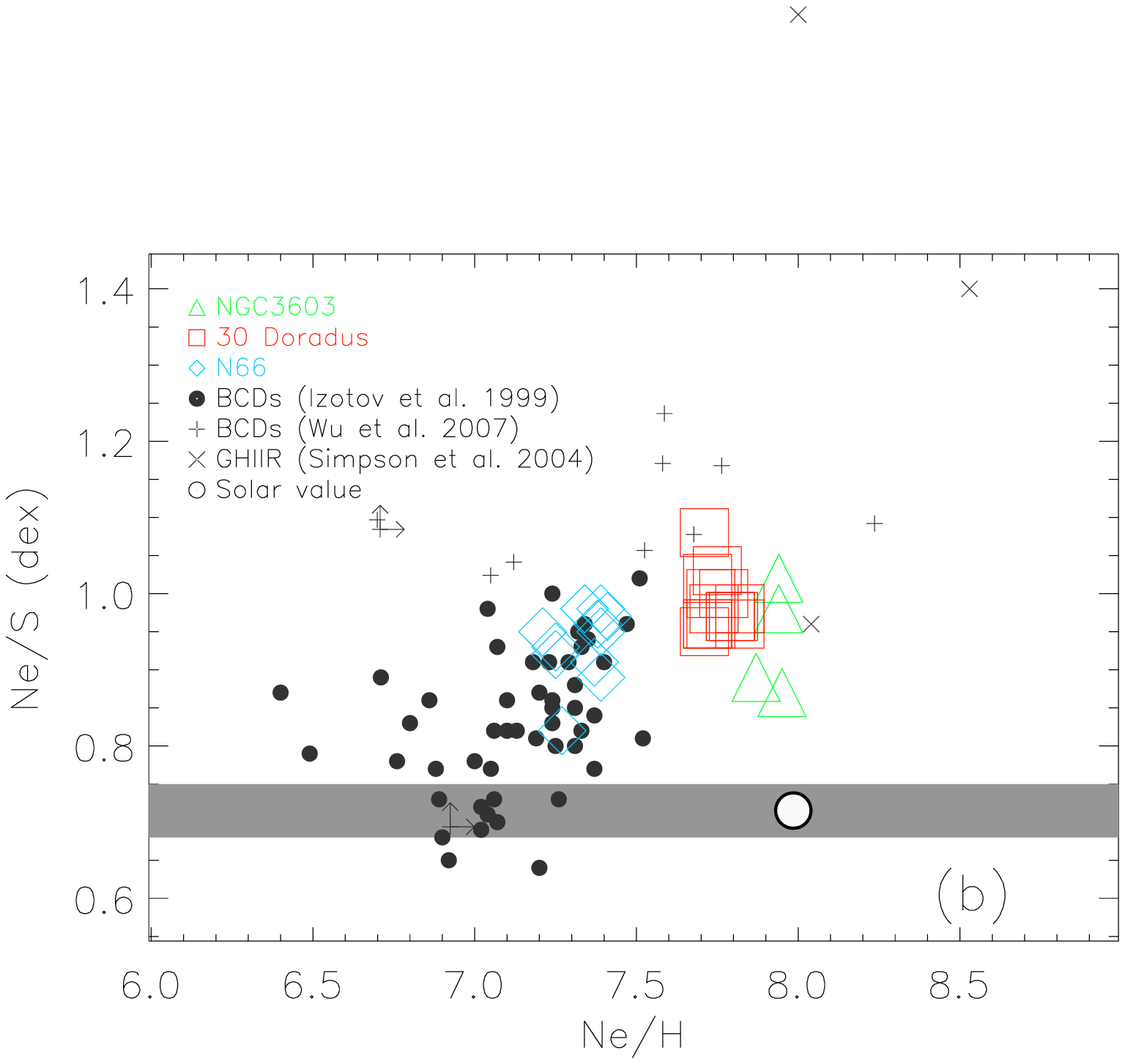}\\
\figcaption{\textit{Top} - S/H is plotted against Ne/H. 'BCDs' stands for blue compact dwarfs, while 'GHIIR' stands for giant H\2\ regions. \textit{Bottom} - The ratio Ne/S is plotted as a function of Ne/H. The gray stripe illustrates the solar proportion between Ne and S, based on the range of the latest solar abundance determinations. The black circle shows the solar abundance ratio. \label{fig:abs}}
\end{figure}

In order to study in more detail the sulfur abundance in the H\2\ regions, we compare it to the neon abundance (see \S\ref{sec:neondisc}).
As discussed in Thuan et al.\ (1995), Ne, S, and Ar are products of $\alpha$-processes during both hydrostatic and explosive 
nucleosynthesis in massive stars. These elements are thus thought to follow a parallel chemical evolution in stars except in the case of an extreme initial mass function (IMF) because the stars synthetizing Ne have somewhat larger masses than the stars producing S and Ar (Woosley \& Weaver 1995).
We observe that Ne/H and S/H trace each other remarkably well in the three objects of our sample (Fig.~\ref{fig:abs}). This contrasts with the lack of correlation found by Verma et al.\ (2003) in starburst regions using ISO data, but this is consistent with
Ne and S production in massive stars.

The Ne/S ratio appears to be significantly larger than the solar value by $\sim$0.2-0.3\,dex (Fig.~\ref{fig:abs}).
Correcting for the presence of S\2\ cannot explain this discrepancy (\S\ref{sec:sulfurdet}). Note that an extreme IMF is also unlikely to explain the sulfur depletion because S/H and 
Ar/H show somewhat different behavior with respect to Ne/H although S and Ar are produced in similar mass stars (\S\ref{sec:argondisc}).

Interestingly, the Ne/S ratio was found to be on average larger in the PNe of the Magellanic Clouds than in the Milky Way, which was attributed to sulfur depletion on molecules and dust grains (Marigo et al.\ 2003; Henry et al.\ 2004; Pottasch \& Bernard-Salas 2006; Bernard-Salas et al.\ 2007). This underabundance of sulfur with respect to other $\alpha$-elements has also been observed in H\2\ regions in the Milky Way (Simpson et al.\ 2004) and in M83 and M33 (Rubin et al.\ 2007) with the same interpretation.  
Unlike Si or Fe for instance, sulfur is not depleted toward cool diffuse clouds (Savage \& Sembach 1996). 
However, in dense regions and in PDRs, it becomes significantly depleted (Simpson \& Rubin 1990; Verma et al.\ 2003; Pottasch \& Bernard-Salas 2006). Goicoechea et al.\ (2006) found that sulfur depletion could be up to a factor 4, i.e., 0.6\,dex in the Horsehead PDR. 
For even denser regions, it has been established that sulfur is about two orders of magnitude more depleted than C, N, and O in molecular clumps with densities $n$(H)$\sim10^{3-5}$\,cm$^{-3}$ (Ruffle et al.\ 1999). 

The position disagreeing the most with the Ne/S solar value is 30DOR\#8, which shows prominent silicate absorption, where the ISM is denser and where refractory elements are more likely to be depleted onto dust grains. 
However, the ionized gas we measure the abundances from is not associated with the dense region, as indicated by the negligible
amount of MIR extinction toward 30DOR\#8 (Lebouteiller et al.\ in preparation). 
Furthermore, regions showing ratios closer to the solar Ne/S (e.g., N66\#7, NGC3603\#3) show PAH emission, which is a sign of relatively dense regions. Hence, though sulfur depletion onto dust grains is likely to occur, there is no trend with the spectral charateristics (PAH-dominated, ionized gas, etc...). This is due to the fact that the ionized gas we probe is not entirely associated with the PDRs or the embedded regions, because of the complex geometry and structure of the lines of sight. 

Results in blue compact dwarfs (Izotov \& Thuan 1999; Wu et al.\ 2007), and Galactic H\2\ regions (Simpson et al.\ 2004) are added for comparison in Figure~\ref{fig:abs}. 
BCDs have optical spectra very similar to Galactic H\2\ regions, and they sample the low-metallicity zone in the plot. 
The giant H\2\ region N\,66 in the SMC overlaps the most metal-rich BCDs.
Note that abundance measurements from the optical and MIR studies of BCDs disagree with each other (crosses and filled circles in Figure~\ref{fig:abs}). However in the case of the BCDs, the difference comes from a higher neon abundance measured in the MIR, which could be due to the fact that the gas probed in the MIR is enriched in heavy elements and that sulfur is depleted on dust grains.  
It seems that the trend formed by the BCDs and the H\2\ regions of our sample is significantly off the solar proportion for metallicities close to solar, while it barely follows the solar proportion at metalllicities lower or equal to that of N\,66. 
Globally, there might be a hint that the Ne/S ratio increases together with the metallicity, which would be an important proof 
of sulfur depletion on dust grains. 


\subsubsection{Argon}\label{sec:argondisc}

\begin{figure}[h!]
\includegraphics[angle=0,scale=0.45,clip=true]{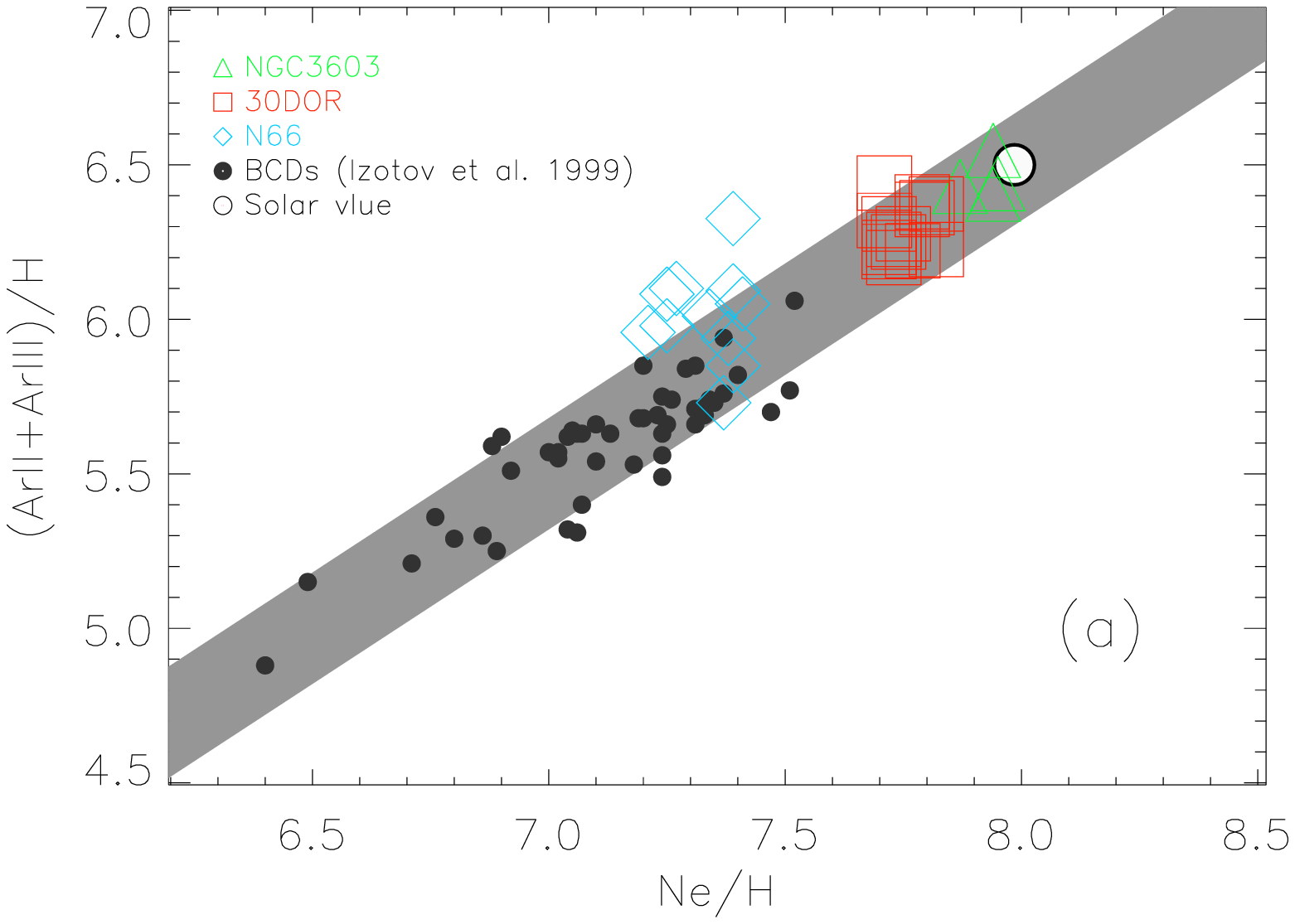}\\
\includegraphics[angle=0,scale=0.45,clip=true]{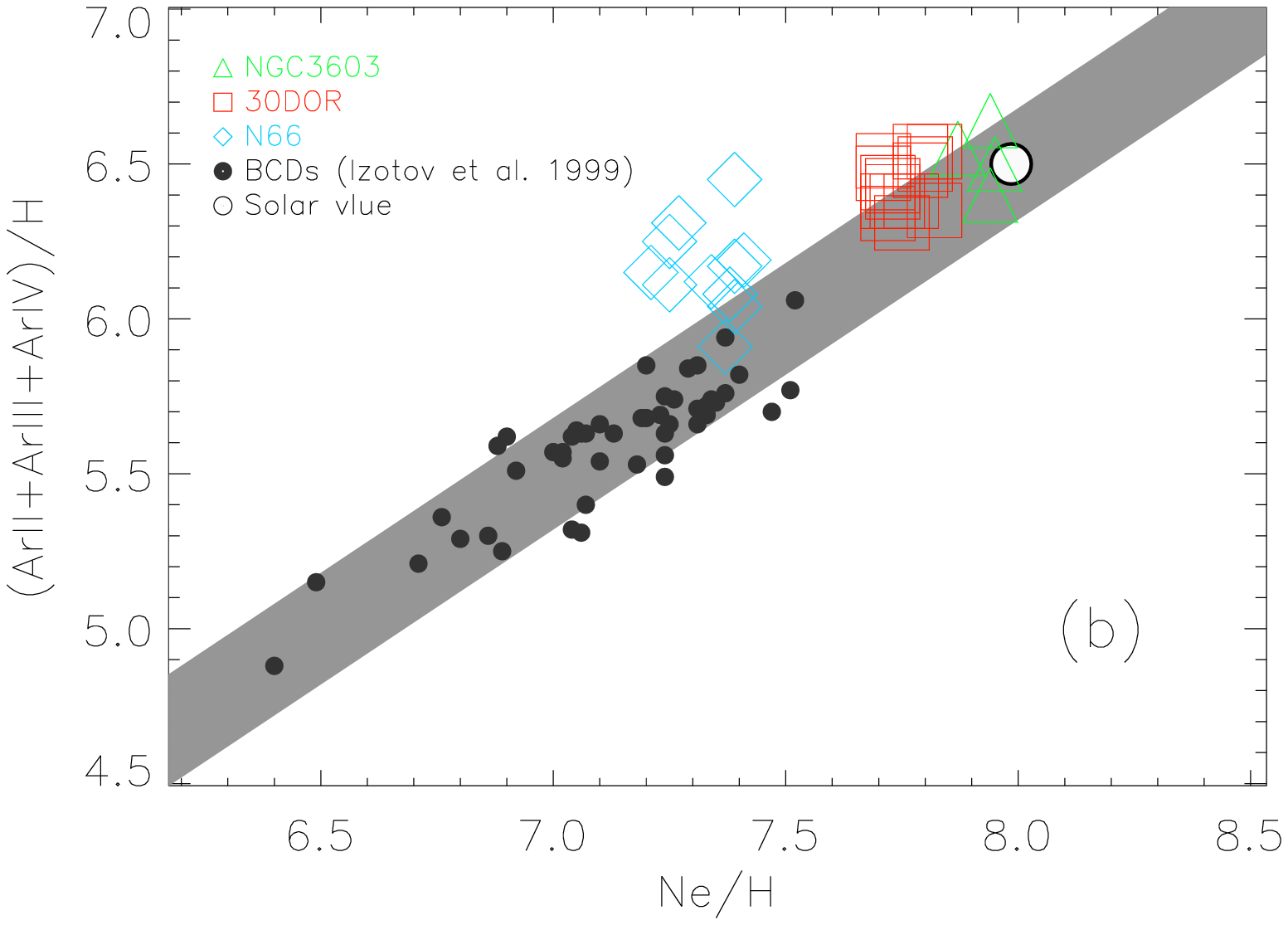}\\
\figcaption{Ar/H is plotted against Ne/H. \textit{top} - Ar\3\ ion contribution is solely used, without ionization corrections. \textit{Bottom} - An ionization correction was applied, due to the presence of Ar\4\ (see \S\ref{sec:argondet}). See Fig.~\ref{fig:abs} for a plot description. \label{fig:abar}}
\end{figure}

In NGC\,3603, Ar/H it is somewhat smaller than the optical determination (Fig.~\ref{fig:abundances}).
The argon abundances in N\,66 and 30\,Dor are surprisingly large even when no ICF accounting for Ar\4\ is applied. 
The argon abundances determined using only the Ar\2\ and Ar\3\ contributions match better the solar and BCD values than with ICF correction due to the possible presence of Ar\4\ (Fig.~\ref{fig:abar}).  
As an illustration, when using an an ICF to account for Ar\4, we find Ar/H in 30\,Dor equal to the highest solar determinations (while 30\,Dor has clearly a subsolar metallicity with 0.60\,Z$_\odot$; \S\ref{sec:cosmic}). 
These results are consistent with the fact that the main ionization stage in the dense gas probed in the MIR is Ar\3, with a small contribution due to Ar\2\ (\S\ref{sec:argondet}).

Argon is produced by oxygen-burning in massive stars, and it is expected to follow both sulfur and neon nucleosynthesis.
Indeed the argon abundance Ar/H correlates with Ne/H in the giant H\2\ regions (Fig.~\ref{fig:abar}), but with a somewhat larger dispersion than the Ne/H vs.\ S/H correlation. This is mostly due to large measurement uncertainties on the [Ar\2] and [Ar\3] line fluxes (\S\ref{sec:measurements}), and not to the contribution from Ar\4\ (\S\ref{sec:argondet}).
There is no position in any of the H\2\ regions showing an argon deficiency with respect to neon, even at high metallicities, which is consistent with argon being undepleted on dust grains or in molecules, even in the densest regions probed.

The MIR neon and argon abundance determinations in NGC\,3603 and 30\,Dor agree well with the solar value (Fig.~\ref{fig:abar}). In contrast, about half of the positions in N\,66 shows a larger argon abundance than expected from the solar Ne/Ar. 
Verma et al.\ (2003) find that argon is overabundant relative to neon in a wide variety of star-forming galaxies, with metallicities from sub-solar (BCDs) to super-solar (WR galaxies). There is no trend among the galaxies in their sample. We could see the effect of enhanced argon abundance in N\,66 but it is more likely due to uncertaities in the abundance determination.

\subsubsection{Iron}\label{sec:irondisc}

\begin{figure}[h!]
\includegraphics[angle=0,scale=0.45,clip=true]{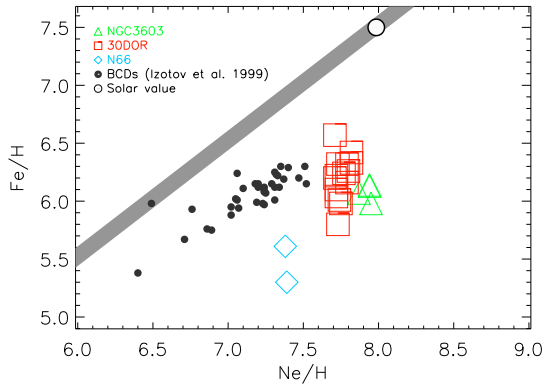}\\
\figcaption{Fe/H is plotted against Ne/H. See Fig.~\ref{fig:abs} for a plot description. \label{fig:abfe}}
\end{figure}

The average Fe/H determinations in NGC\,3603 and 30\,Dor agree with each other (Table~\ref{tab:eleab}), 
and barely match Fe/H in the most metal-rich BCDs (Fig.~\ref{fig:abfe}). 
The average iron abundance in NGC\,3603 ($6.09\pm0.37\pm0.08$) compares well with the value derived from the optical ($6.05\pm0.10$; Garc{\'{\i}}a-Rojas et al.\ 2006), and implies no differential depletion between the gases probed in the MIR and in the optical (\S\ref{sec:diffdep}).

The iron abundance in 30\,Dor shows a large dispersion including the optical value within the range.
The Fe/H value toward 30DOR\#17 is more than 3$\sigma$ larger than the average iron abundance in 30\,Dor. 
The spectrum of 30DOR\#17 shows deep silicate absorption probably originating from dust associated with an ultracompact H\2\ region.
In the same spectrum, we also observe the high-excitation [O\4] line at 25.89\mic, which is known to originate around hot Wolf-Rayet stars (Schaerer \& Stasinska 1999)
and in shock-heated gas (Lutz et al.\ 1998). Because the region is dominated by schocks from the nearby supernova remnant (SNR) 30\,Dor\,B (Chu et al.\ 1992), it is likely that the [O\4] line traces gas shocked by the SN. 
Hence the enhanced iron abundance toward 30DOR\#17 is consistent with removal of iron atoms from dust grains due to shocks.
We plan to investigate in more detail the ISM properties toward the SNR using the new data provided by an accepted IRS Cycle 5 GTO program.

There are only two Fe/H determinations in N\,66 (positions \#1 and \#8), both of which are significantly smaller than the Fe/H values in the two other giant H\2\ regions, and also smaller than the BCDs having similar metallicities (as traced by Ne/H). This might be due to underestimated ionization corrections due to the presence of Fe\4\ in this object (\S\ref{sec:irondet}). However, although N\,66 is characterized by a globally harder ISRF as compared to the two other H\2\ regions in our sample, the positions N66\#1 and N66\#8 do not show significantly harder ISRFs than all the positions in 30\,Dor. Hence the iron underabundance in N\,66 could be genuine.

The Fe/Ne ratio in the three giant H\2\ regions is significantly lower than the solar value (Fig.~\ref{fig:abfe}). This can be explained by iron depletion on the surface of very small grains, which are the dominant dust component in the ionized gas
(see e.g., Cesarsky et al.\ 1996; Verstaete et al.\ 1996; Lebouteiller et al.\ 2007). 
Rodr{\'i}guez (2002) observed that the [O/Fe] ratio increases with metallicity in BCDs, and explained it by iron depletion on dust at high metallicity. We find that the trend formed by the BCDs and the giant H\2\ regions flattens for large neon abundances, suggesting that the iron becomes more and more underabundant as the metallicity increases, being consistent with iron depletion on dust grains.

\subsection{Dense gas properties}\label{sec:abopt}

\subsubsection{Depletion onto dust grains}\label{sec:diffdep}

MIR spectra allow us to probe regions that are obscured at optical wavelengths, possibly with different physical and chemical properties. 
The average sulfur abundances we calculated from IRS spectral lines in the giant H\2\ regions corroborate the optical determinations (Fig.~\ref{fig:abundances}), even though they likely fall towards their lower side. 
Interestingly, Simpson et al.\ (1995) measured the MIR [S\3] line flux, and used the [S\4] line flux in the literature to infer S/H=7.12 in NGC\,3603, i.e., close 
to our average determination for this object ($6.99\pm0.11\pm0.08$; Table~\ref{tab:eleab}). These two MIR determinations are significantly lower than 7.36$\pm$0.08, which was obtained from optical lines by Garc{\'{\i}}a-Rojas et al.\ (2006). 
If the relative sulfur underabundance in the MIR is real, this could hint at a differential sulfur depletion between the gases probed in the MIR and in the optical. However, it is likely that the underabundance is driven by uncertainties in the abundance determination methods.
Ionization correction in the MIR could represent 0.04-0.05\,dex (\S\ref{sec:sulfurdet}). 
On the other hand, optical studies do not directly observe S\4, which in our case contribute around 5\% to the total sulfur in NGC\,3603, and as much as 24\% in N\,66 and 17\% in 30\,Dor. The ICF applied in the optical could overcorrect the total sulfur abundance. 
Morevover, electronic temperature variations/uncertainties affect the optical determinations and to a less extent the MIR determinations (\S\ref{sec:ionicab_results}). 




The study of the refractory element iron supports the lack of differential depletion between the gases probed in the MIR and in the optical.
We found that the iron abundance in NGC\,3603 and 30\,Dor is similar when derived in the optical and in the MIR (\S\ref{sec:irondisc}).
This confirms that the possible discrepancy seen in S/H between MIR and optical observations is not due to additional depletion in the gas observed in the MIR, but rather to uncertainties on the abundance determination. 
This stresses the importance of deriving the abundance of a refractory element such as iron in dense regions.

\subsubsection{Ionization degree}\label{sec:iondegree}

We found that the chemical composition derived in the optical and in the MIR is overall fairly similar (Fig.~\ref{fig:abundances}). For this reason, it is possible to compare the ionic abundances from optical and MIR wavelengths and interpret possible differences in physical conditions. 
In order to trace the gas excitation, we use the ionic abundance ratio (Ne\3/H)/(S\3/H). This ratio is a good approximation for (Ne\3/H)/(Ne\2/H) in H\2\ regions (Lebouteiller et al.\ in preparation) and it can be derived in both optical and MIR ranges. 
Based on the values in Table~\ref{tab:ionicab_global}, the (Ne\3/H)/(S\3/H) ratio is $3.7\pm0.7$ in NGC\,3603, i.e., about two times smaller
than the value in the optical ($7.7\pm0.4$; Garcia-Rojas et al.\ 2006). We find (Ne\3/H)/(S\3/H) = $7.6\pm1.5$ in 30\,Dor, somewhat smaller than
$10.7\pm0.1$ in the optical (Peimbert 2003). Finally, we find (Ne\3/H)/(S\3/H) = $9.8\pm2.0$ in N\,66, as compared to 10.3 in the optical (Vermeij \& van der Hulst 2002). 
Hence the MIR spectra seem to probe ionized gas with a degree of ionization equal to or lower than that of the gas probed in the optical.
This could be due to the fact that the SH and LH apertures from the IRS do not sample the same spatial regions as those observed within narrow slits used in the optical. 
It is also possible that the gases probed in the MIR and in the optical have different physical conditions (such as density).

\subsection{Metal dispersion and mixing}\label{sec:cosmic}

\subsubsection{Metal enrichment history}\label{sec:meh}

Stellar winds and supernov\ae\ explosions from the current star-formation episode release newly produced elements into the surrounding ISM on a short timescale (few $10^{6}$\,yr). It is thus natural to question the origin of the metals we observe in the ionized gas of the giant H\2\ regions. We investigate in this section the total metal content of each giant H\2\ region by comparing their metallicities with the Sun and other objects from the same host galaxies (H\2\ regions, planetary nebul\ae).

First it must be stressed that the solar abundance determinations have shown strong variations over the past few years (see e.g., Pottasch \& Bernard-Salas 2006). 
The gray stripes in Fig.~\ref{fig:abundances} illustrate the range of solar abundance determinations for each element over the 1998-2007 period. Neon and argon abundances are
particularly poorly determined because of the absence of suitable lines in the solar photosphere. Abundance determinations
of those elements are indirect and make use of coronal lines together with a correction using other $\alpha$-elements
such as magnesium or oxygen. Considering the most extreme determinations among the recent studies (Asplund et al.\ 2006; 
Feldman \& Widing 2003), solar Ne/H determinations range from $7.84\pm0.06$ to $8.08\pm0.06$ while Ar/H ranges from $6.18\pm0.06$ to $6.62\pm0.06$.
The sulfur solar abundance is better determined, with values from 7.14 (Asplund et al.\ 2006) to 7.33 (Grevesse et al.\ 1998).

We consider neon as being the most reliable metallicity tracer available in our study, as compared to sulfur, argon, and iron (\S\ref{sec:neondisc}). 
The average neon abundance is $7.94(\pm0.11\pm0.08)$ in NGC\,3603, $7.76(\pm0.11\pm0.08)$ in 30\,Dor, and $7.34(\pm0.11\pm0.08)$ in N\,66 (Table~\ref{tab:eleab}).
Considering an average solar value of 7.98,
the average neon abundance in NGC\,3603 implies essentially a solar metallicity ($\approx$0.91~Z$_\odot$), while 30\,Dor is $\approx$0.60~Z$_\odot$, and N\,66 is $\approx$0.23~Z$_\odot$.

The metallicity of NGC\,3603 agrees with the solar metallicity within uncertainties.
This is consistent with the abundance gradient in the Milky Way (see, e.g., Martin-Hernandez et al.\ 2002) given the fact that the galactocentric distance of NGC\,3603 is almost that of the Sun (8.5\,kpc as compared to 8\,kpc; \S\ref{sec:pres_3603}). It is instructive to compare our abundances in 30\,Dor and N\,66 with those in planetary nebul\ae\ (PNe) and in other H\2\ regions in the LMC and SMC from Bernard-Salas et al.\ (2007). The authors measured abundances using MIR lines in a similar way to the present study so that the comparison does not suffer from significant systematic uncertainties. Bernard-Salas et al.\ (2007) found that neon and sulfur abundances in PNe agree relatively well with those in the H\2\ regions of the LMC and SMC, implying that H\2\ regions have not been significantly enriched over the past few Gyrs (age of the PNe progenitor). The global neon abundance we inferred from the MIR lines in 30\,Dor is 
$7.76(\pm0.11\pm0.08)$, remarkably close to the average value in the PNe of the LMC $<$Ne/H$>$=7.78 (no quoted errors).
The global neon abundance we derive in N\,66 ($7.34\pm0.11\pm0.08$) is consistent within error bars with the average $<$Ne/H$>$ in the PNe of the SMC (7.43). 
In addition, it must be noticed that the metallicity in O dwarf stars in NGC\,346, the stellar cluster associated with N\,66, is 0.2~Z$_\odot$ (Bouret et al.\ 2003), which agrees well with our metallicity determination in the ISM of the H\2\ region ($\approx$0.23~Z$_\odot$). 

It seems that the chemical composition of giant H\2\ regions compares well with that of PNe and young stars, suggesting that the ionized gas of the H\2\ regions has not been enriched significantly by the current star-formation episode, or by any previous episode more recent than $\lesssim$1\,Gyr. If enrichment has really occured, the metallicity of the H\2\ regions has been enhanced by much less than a factor 2.
Full mixing requires that elements acquire the same physical conditions as the ionized gas, namely temperature and density, but also viscosity and molecular diffusion (Scalo \& Elmegreen 2004). The effective mixing provided by turbulent diffusion in H\2\ regions allows metals to mix on spatial scales of a few hundreds parsecs over 100\,Myr (Roy \& Kunth 1995; Avillez \& Mac Low 2002). Hence abundance discontinuities between H\2\ regions in a single galaxy should be ubiquitous, especially in low-mass galaxies where rotational shear is weak (Roy \& Kunth 1995). 
However observations challenge this hypothesis and suggest that mixing can occur on spatial-scales as large as the galaxy size and over time-scales larger than the age of the H\2\ regions (see e.g., Kobulnicky 1998; Skillman \& Kennicut 1993; Noeske et al.\ 2000; Russel \& Dopita 1990). 
Our results support such spatial- and time-scales and question the fate of metals released by massive stars in giant H\2\ regions.
Following the scenario of Tenorio-Tagle (1996), the metals could undergo a long cycle in a hot phase before mixing with the surrounding ISM. The composite \textit{IRAC} and \textit{Chandra} image of 30\,Dor shows large cavities filled with hot gas, surrounded by colder and denser ionized gas shells and filaments (Brandl et al.\ in preparation). These are supernova-driven outflows generated by the superstellar cluster (see also e.g., Redman et al.\ 2003). The temperature of metals in the hot cavities is too high for recombination to occur, and the density too small for collisionally-excited levels to be populated so that the abundances we derive from the MIR spectra do not account for these newly released elements. The superbubbles of hot gas in 30\,Dor should remain inside the galaxy for hundreds of Myr according to the scenario of Recchi et al.\ (2001). It could eventually lead to a superwind that will enable the escape of metals in the halo (Tenorio-Tagle 1996).

\subsubsection{Small-scale mixing}\label{sec:dispersion}


Our observations probe the chemical abundances toward several positions within each giant H\2\ region. 
Neon and sulfur abundances show remarkably little dispersion in the three giant H\2\ regions (Fig.~\ref{fig:abundances}). 
Vermeij \& van der Hulst (2002) analyzed 4 positions in 30\,Dor and found neon abundances ranging between 7.84 and 7.94. Analyzing 15 positions, we find a comparable dispersion, 0.11\,dex from the mininum to the maximum Ne/H values. In the same object, Vermeij \& van der Hulst (2002) find sulfur abundances ranging from 6.63 to 6.86, and argon abundances ranging from 6.18 to 6.63, which is a smaller dispersion than we measured (but with fewer positions).
Our results in NGC\,3603 and N\,66 also imply that the neon abundance and, to a lesser extent, the sulfur abundance show little dispersion in these regions. 
Part, if not all, of the dispersion of the elemental abundances is due to uncertainties in the abundance determinations, such as ionization corrections (\S\ref{sec:eleab}) and the assumed nebula physical conditions (\S\ref{sec:ionicab_results}). In this section, we investigate whether there is evidence for small-scale mixing.

The apparent homogeneity of the abundances contrasts with the wide variety of physical regions each line of sight samples toward each object (PDR, stellar cluster, ionized gas, embedded regions). This suggests that the ionized gas from which we measure the chemical abundances is not necessarily associated with these physical regions. In fact, the lack of MIR line extinction toward PDRs and toward embedded objects implies that the whole nebula is filled with ionized foreground gas (Lebouteiller et al.\ in preparation).
Can abundance inhomogeneities still be inferred? 
The material ejected from supernov\ae\ could reach a hot coronal phase before falling back onto the galactic disk in the form of molecular droplets (Tenorio-Tagle 1996). This material will then be photodissociated by massive stars and mix with the surrounding ionized gas. According to this model, mixing could occur at very small spatial scales ($\lesssim$1\,pc). Tsamis \& Pequignot (2005) introduced in their model of 30\,Dor small-scale chemical inhomogeneities at the subparsec-sized scale. The typical distance between the IRS observations is $\sim$4\,pc in NGC\,3603, $\sim$20\,pc in 30\,Dor, and $\sim$15\,pc in N\,66. Hence our MIR observations do not have the necessary spatial resolution to probe such chemical inhomogeneities.

On the other hand, Recchi et al.\ (2001) proposed that, mostly because of thermal conduction, the SNe type II ejecta start to cool down and mix with the cold gas within a few $10^{6}$\,yr before the formation of hot gas outflows. In this particular case, small-scale mixing could be observed locally. 
In order to assert whether small-scale mixing is responsible for the abundance variations we observe across each giant H\2\ region, we use Ne/H in 30\,Dor as a reference.
The standard deviation of Ne/H is 0.02\,dex, implying that the abundance of neon could vary as much as $\approx$5\% 
across the region. Interestingly, 5\% of the neon abundance currently present in 30\,Dor corresponds to the total neon enrichment that the 
star-forming dwarf galaxy IZw18 experienced so far (Ne/H = 6.40; Izotov et al.\ 1999). Kunth \& Sargent (1986) proposed that the metallicity of IZw18 represents the minimum chemical enrichment of an H\2\ region from a starburst episode. It is unlikely that the small-scale variations in a single giant H\2\ region correspond to the yields from a starburst episode, especially considering the fact that the metals in the ionized gas are not likely to be cotemporal with the star-formation episode (\S\ref{sec:cosmic}).
Uncertainties on the measurements and on the abundance determination method ($\pm0.11\pm0.08$\,dex for Ne/H, i.e., as much as 55\%) are too large to constrain possible abundance fluctuations of less than 5\% at the metallicity of 30\,Doradus. 

\section{Conclusions}

We analyzed the chemical abundances in the ISM of three giant H\2\ regions, NGC\,3603 in the Milky Way, 30\,Dor in the LMC, and N\,66 in the SMC using the MIR lines observed with the IRS onboard Spitzer.
\begin{itemize}

\item Our observations probe the ISM toward various physical regions, such as stellar clusters, ionized gas, photodissociation regions, and deeply embedded MIR bright sources. The spectra show the main ionization stages of neon, sulfur, and argon in the ionized gas. We also detect [Fe\2] and [Fe\3] lines.

\item Ionic abundances of Ne\2, Ne\3, S\3, S\4, Ar\2, Ar\3, Fe\2, and Fe\3\ were derived. The internal variation of electron density across a region has no impact on the ionic abundance determination. On the other hand, we find that electron temperature uncertainties and/or intrinsic variations could be responsible for an error of 20\% at most on the abundance determinations. Based on the (Ne\3/H)/(S\3/H) ionic abundance ratio, we find that the optical spectra probe a gas with a degree of ionization equal to or higher than the gas probed in the MIR.

\item Elemental abundances were determined from the ionic abundances. No ionization corrections were needed, except for iron. We find that neon, sulfur, and argon scale with each other, which is expected from stellar yields. Abundances do not show any dependence on the physical region (PDR, stellar cluster, embedded region, ...).

\item The Ne/S ratio is larger than the solar value, and suggests that sulfur could be depleted onto dust grains. The sulfur abundance in the MIR agrees best with the lowest optical determinations, which is likely due to uncertainties in the abundance determinations.

\item Iron abundance shows a larger uncertainty than Ne/H, S/H, and Ar/H. The comparison of iron and neon abundances hints at significant depletion of iron onto dust grains at large metallicities. The agreement with the optical determination of Fe/H indicates however that there is no differential depletion on dust grains between the gas probed in the MIR and in the optical. 

\item  Fe/H is found to be spectacularly large in one position, corresponding to a supernova remnant. This strongly suggest that iron atoms have been released from dust grains due to schocks from the SN.

\item The metallicity of NGC\,3603 agrees with the Galactic abundance gradient. The metallicities of 30\,Doradus and N\,66 agree well with those of the PNe in their respective host galaxies. These findings suggest that the giant H\2\ regions did not experience a significant metal enrichment for at least 1\,Gyr. If enrichement occured, the metallicity was altered by less than a factor of two.

\item Neon and sulfur abundances show remarkably little dispersion in the three H\2\ regions (e.g., 0.11\,dex dispersion in 15 positions in 30\,Dor). Small-scale mixing is apparently effective, abundance fluctuations are smaller than $\sim$55\%.
However, internal variations of the abundances are likely to be on the order of $\lesssim$5\%, and determining their existence would require a 
significant improvement of the data quality and of the method to be evidenced.
\end{itemize}

\acknowledgments We would like to thank Daniel Kunth, Suzanne Madden, and Marc Sauvage for their useful comments. We thank the referee for providing constructive comments. This work is based on observations made with the {\it Spitzer Space Telescope}, which is operated
by the Jet Propulsion Laboratory, California Institute of Technology, under NASA contract
1047. Support for this work was provided by NASA through contract 1257184 issued by JPL/Caltech. VC would like to acknowledge the partial support
from the EU ToK grant 39965.


\appendix

\section{Glossary of abbreviations}

\textit{BCD} : Blue compact dwarf.

\textit{CEL} : Collisionally-excited line.

\textit{GTO} : Guaranted time observation.

\textit{ICF} : Ionization correction factor.

\textit{IMF} : Initial mass function.

\textit{IP} : Ionization potential.

\textit{ISM} : Interstellar medium.

\textit{ISRF} : Interstellar radiation field.

\textit{LMC} : Large Magellanic Cloud.

\textit{PAH} : Polycyclic aromatic hydrocarbon.

\textit{PDR} : Photodissociation region.

\textit{PICS} : Photoionization cross-section.

\textit{SMC} : Small Magellanic Cloud.

\textit{SNR} : Supernova remnant.

\section{Spectra}

\begin{figure*}
\includegraphics[angle=0,scale=1.0,clip=true]{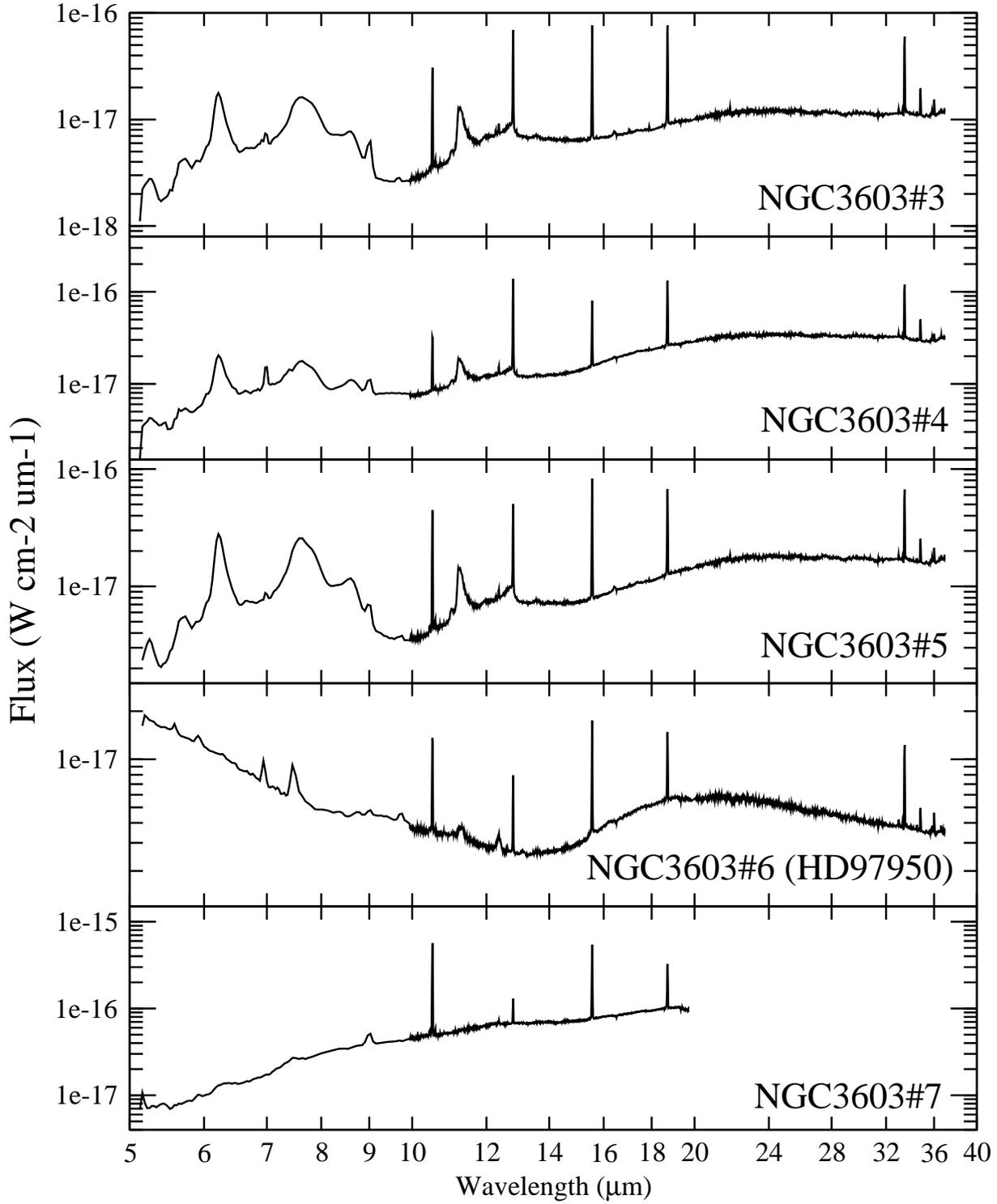}
\figcaption{Spectra of the MIR sources in NGC\,3603 (Table~\ref{tab:observations}). The spectral window 20-36\mic\ in LH is not shown in positions \#7, \#8, and \#9 because of saturated lines and continuum.
\label{fig:spectra_3603}}
\end{figure*}

\begin{figure*}
\includegraphics[angle=0,scale=1.0,clip=true]{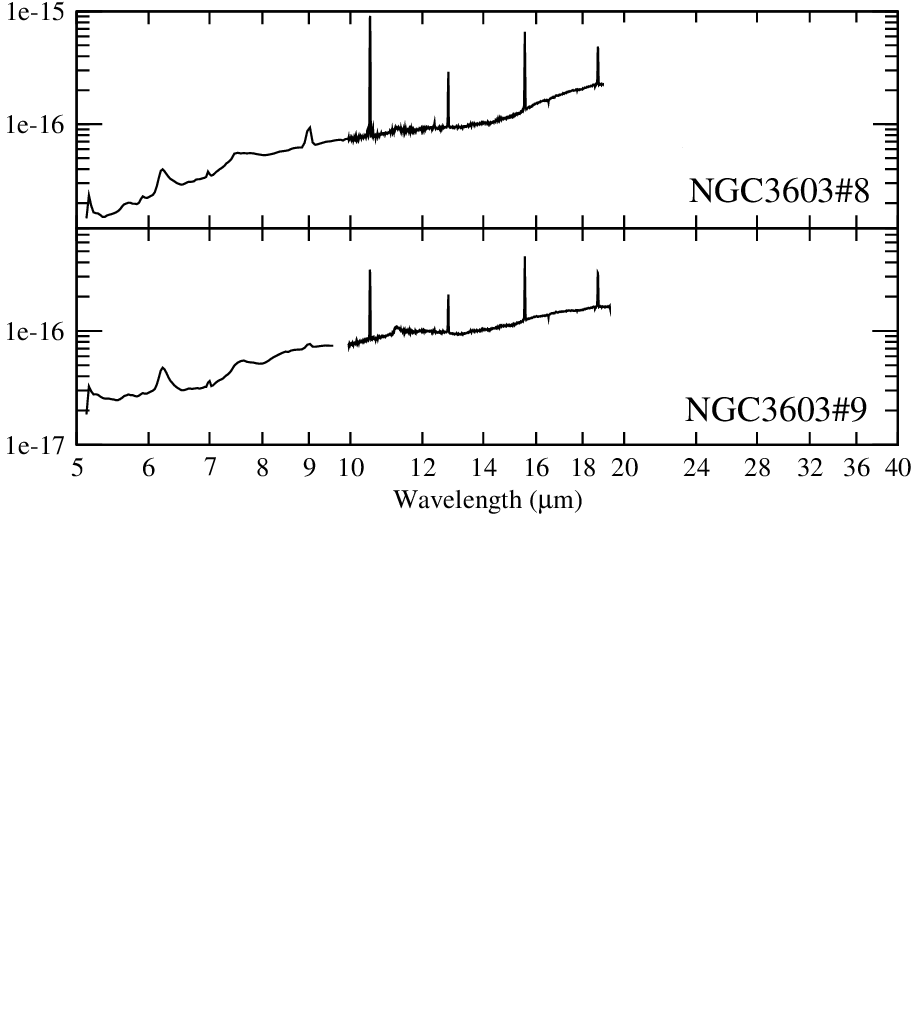}
\figcaption{Spectra of the MIR sources in NGC\,3603 (Table~\ref{tab:observations}). The spectral window 20-36\mic\ in LH is not shown in positions \#7, \#8, and \#9 because of saturated lines and continuum.
\label{fig:spectra_36032}}
\end{figure*}

\begin{figure*}[h!]
\includegraphics[angle=0,scale=1.0,clip=true]{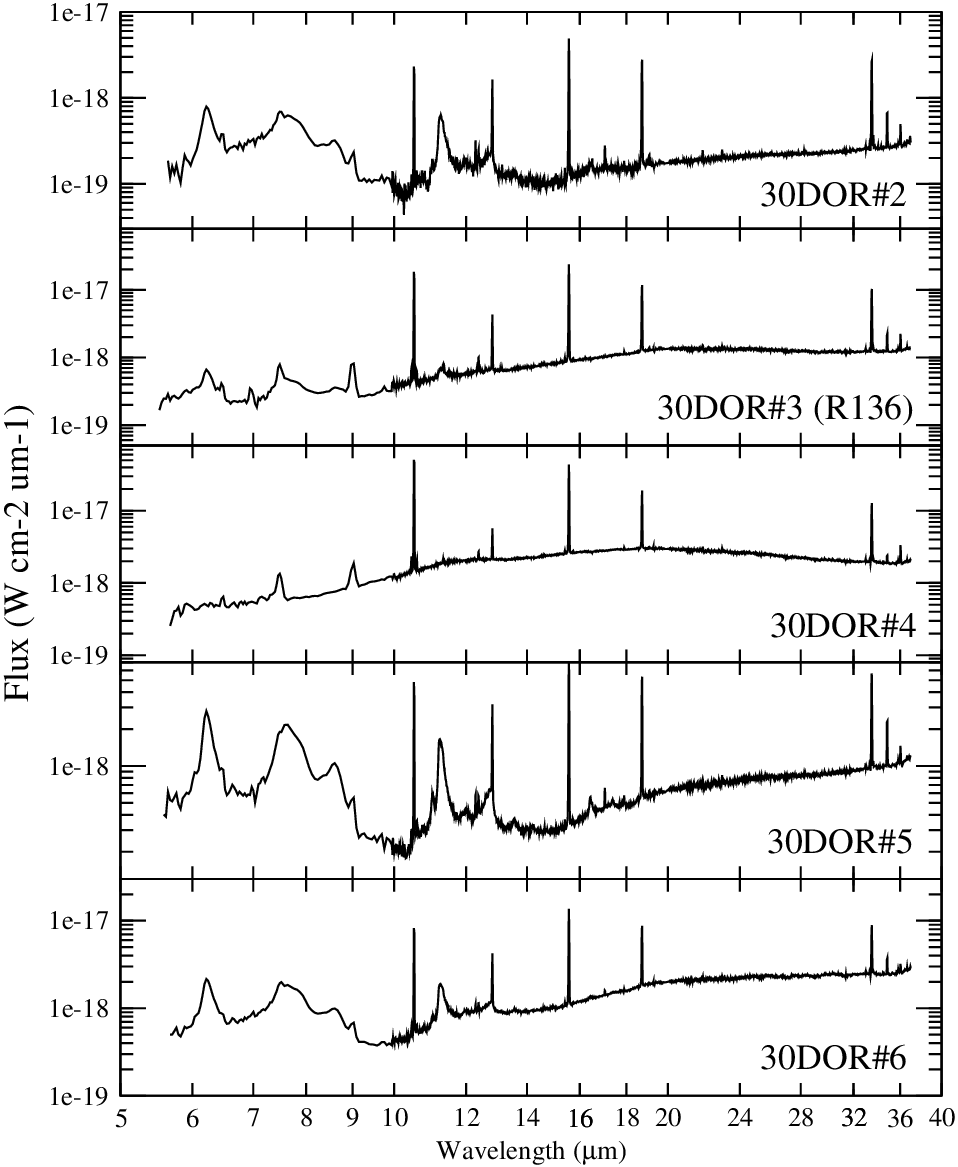}
\figcaption{Spectra of MIR sources in 30\,Dor (Table~\ref{tab:observations}).
\label{fig:spectra_30}}
\end{figure*}

\begin{figure*}[h!]
\includegraphics[angle=0,scale=1.0,clip=true]{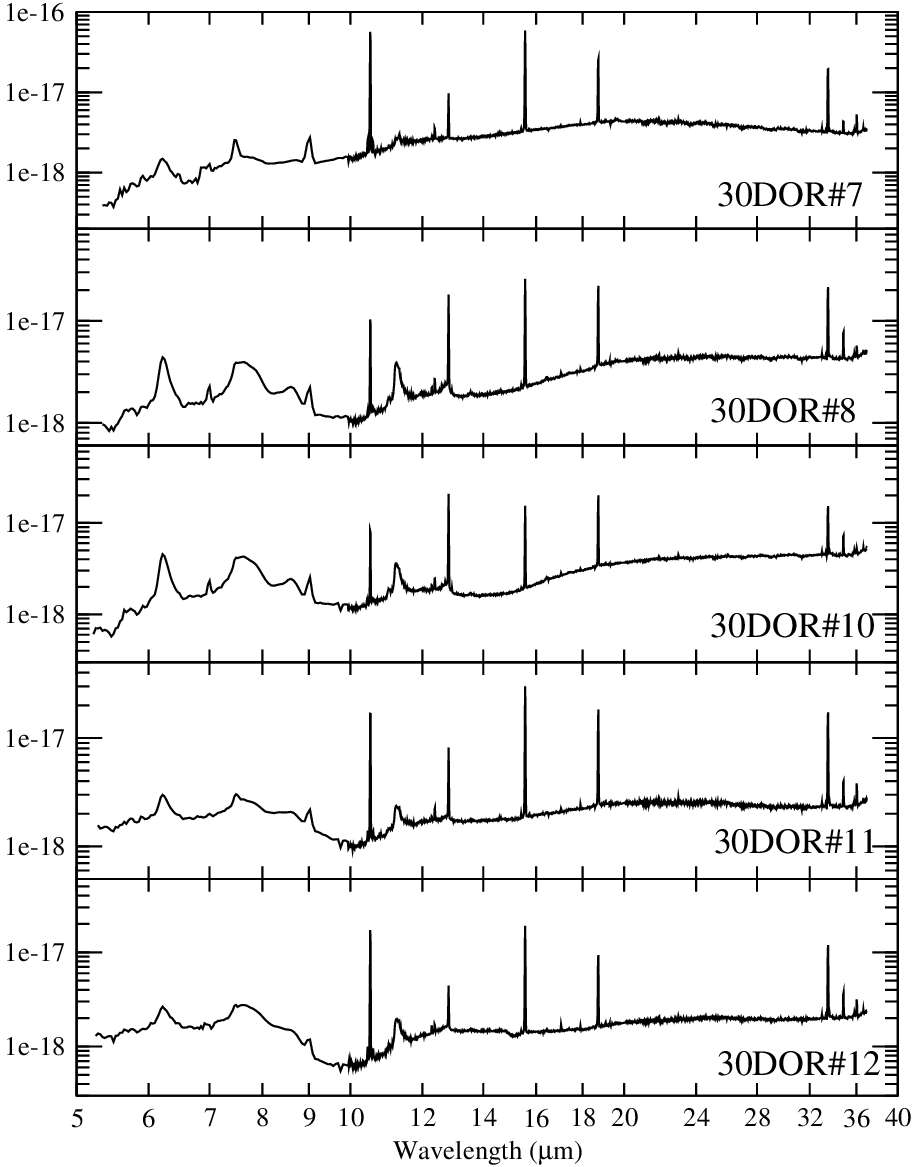}
\figcaption{Spectra of MIR sources in 30\,Dor (Table~\ref{tab:observations}).
\label{fig:spectra_302}}
\end{figure*}

\begin{figure*}[h!]
\includegraphics[angle=0,scale=1.0,clip=true]{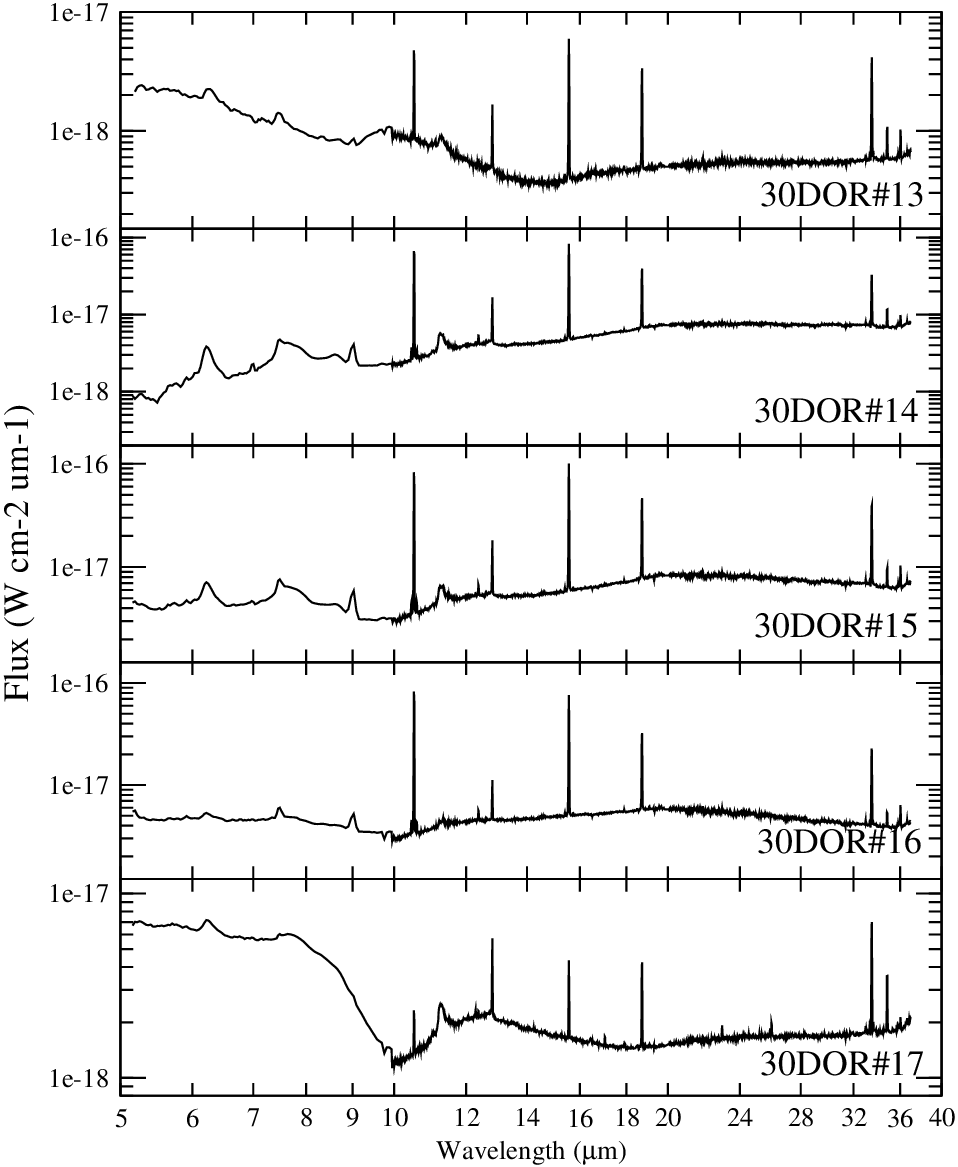}
\figcaption{Spectra of MIR sources in 30\,Dor (Table~\ref{tab:observations}).
\label{fig:spectra_303}}
\end{figure*}

\begin{figure*}[h!]
\includegraphics[angle=0,scale=1.0,clip=true]{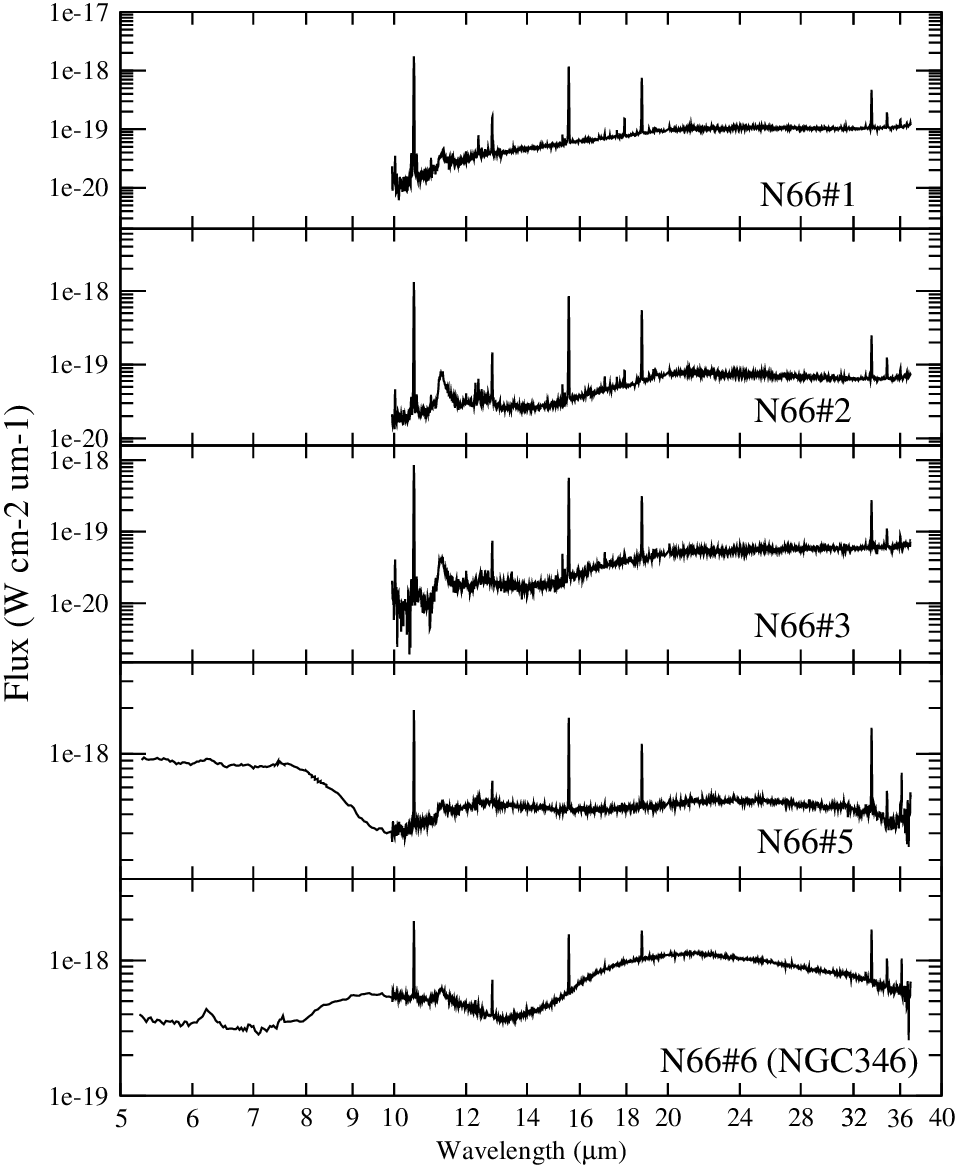}
\figcaption{Spectra of MIR sources in N\,66 (Table~\ref{tab:observations}). Posisions \#1, \#2, and \#3 were observed only with the high-resolution modules.
\label{fig:spectra_346}}
\end{figure*}

\begin{figure*}[h!]
\includegraphics[angle=0,scale=1.0,clip=true]{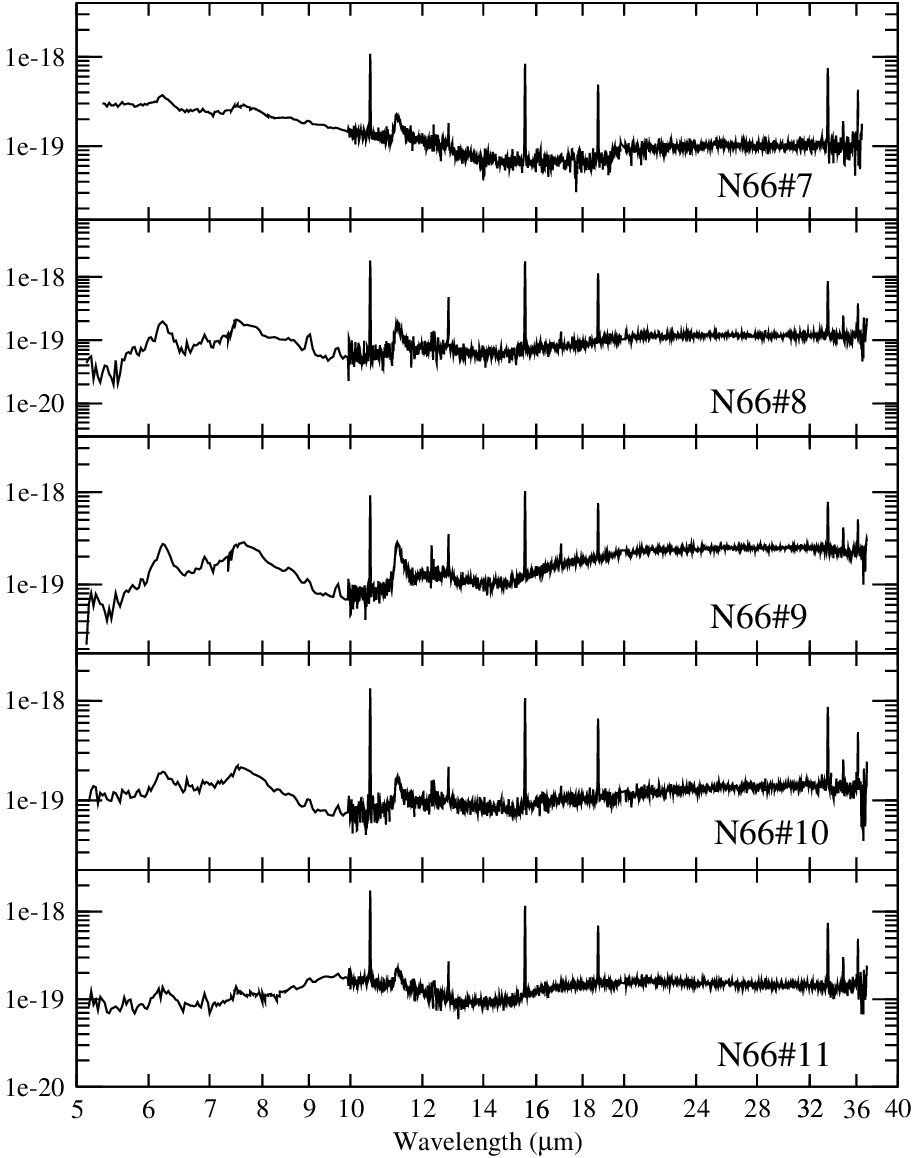}
\figcaption{Spectra of MIR sources in N\,66 (Table~\ref{tab:observations}). Posisions \#1, \#2, and \#3 were observed only with the high-resolution modules.
\label{fig:spectra_346}}
\end{figure*}

\begin{figure*}[h!]
\includegraphics[angle=0,scale=1.0,clip=true]{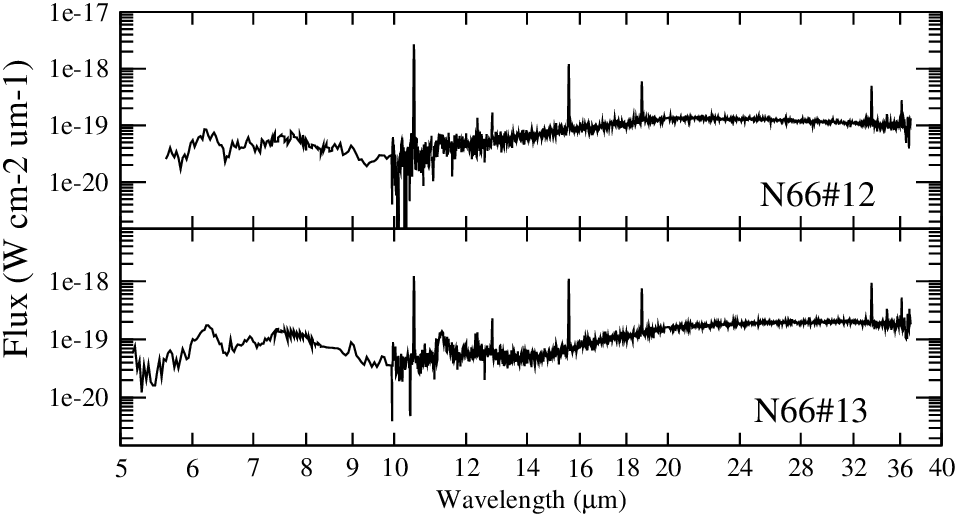}
\figcaption{Spectra of MIR sources in N\,66 (Table~\ref{tab:observations}). Posisions \#1, \#2, and \#3 were observed only with the high-resolution modules.
\label{fig:spectra_346}}
\end{figure*}

\end{document}